# A general approach for the synthesis of two-dimensional binary compounds

Abhay Shivayogimath[1,2], Joachim Dahl Thomsen[1,2], David M. A. Mackenzie[1,2], Mathias Geisler[2,3], Jens Kling[4], Zoltan Imre Balogh[4], Andrea Crovetto[1,5], Patrick R. Whelan[1,2], Peter Bøggild[1,2], Timothy J. Booth[1,2]*

[1]*DTU Nanotech, Technical University of Denmark, Ørsteds Plads 345E, DK-2800 Kgs. Lyngby, Denmark*

[2]*Centre for Nanostructured Graphene (CNG), Technical University of Denmark, Ørsteds Plads 345C, DK-2800 Kgs. Lyngby, Denmark*

[3]*DTU Fotonik, Technical University of Denmark, Ørsteds Plads 343, DK-2800 Kgs. Lyngby, Denmark*

[4]*DTU Cen (Center for Electron Nanoscopy), Technical University of Denmark, Fysikvej 307, DK-2800 Kgs. Lyngby, Denmark*

[5]*V-SUSTAIN, Villum Center for the Science of Sustainable Fuels and Chemicals, Technical University of Denmark, DK-2800 Kgs. Lyngby, Denmark*

*email: tim.booth@nanotech.dtu.dk

## Abstract

Only a few of the vast range of potential two-dimensional materials have been isolated or synthesised to date. Typically, 2D materials are discovered by mechanically exfoliating naturally occurring bulk crystals to produce atomically thin layers, after which a material-specific vapour synthesis method must be developed to grow interesting candidates in a scalable manner. Here we show a general approach for synthesising thin layers of two-dimensional binary compounds. We apply the method to obtain high quality, epitaxial $MoS_2$ films, and extend the principle to the synthesis of a wide range of other materials - both well-known and never-before isolated - including transition metal sulphides, selenides, tellurides, and nitrides. This approach greatly simplifies the synthesis of currently known materials, and provides a general framework for synthesising both predicted and unexpected new 2D compounds.

## Introduction

A large proportion of the possible 2D materials are binary compounds with the designation $MX_n$, where M is typically a transition metal and X a chalcogen or non-metal from groups IV, V, and VI[1–5]. The molybdenum and tungsten disulphides and diselenides remain the most commonly studied 2D binary compounds - with the notable exception of hexagonal boron nitride (hBN) - due to the ready availability of naturally occurring bulk crystals amenable to exfoliation. Chemical vapour deposition (CVD) techniques for the scalable synthesis of these materials are available[6]: however, controlling the stoichiometry and hence the defect density can be challenging. Such techniques typically employ solid metal oxide[7–10] or metal-organic[11] precursors which are chalcogenated at elevated temperatures.

Finding appropriate metal precursors can be a limiting challenge for extending these methods to other 2D transition metal compounds, and the methods typically require dedicated processes and equipment that are highly optimised for growing one specific material.

Here we present a general method for synthesising two-dimensional compounds from elemental solid metal precursors (figure 1). Published CVD growth models for binary compounds stipulate that both elements be insoluble in the catalyst to ensure surface-limited growth, by analogy with CVD graphene growth on copper. In fact, only one component of a binary compound need be insoluble to achieve surface-limited growth, as demonstrated in a recent in-situ X-ray photoelectron spectroscopy (XPS) study of hBN CVD on copper[12]. In that study, monoatomic layers are grown on the catalyst surface despite the solubility of boron in copper.

In the present work, we emulate similar conditions by alloying metal M films with gold, which has limited solubility of the X elements (X = S, Se, Te, N). In brief, a thin layer (~ 20 nm) of metal M is sputtered onto a *c*-plane sapphire substrate and then sputter coated with a thick layer (~ 500 nm) of gold (see Methods). The M-Au layer is then heated to 850°C to form an alloy in the reaction chamber. The relative thicknesses of the M and Au layers determines the concentration of M in the final alloy, which here is deliberately limited to ≤ 5 at. % in order to maintain single-phase alloying conditions. The Au-M alloy is subsequently exposed to a vapour-phase precursor of element X. The limited solubility of X in the gold restricts the formation of $MX_n$ compounds to the surface of the alloy, resulting in few-atom thick layers of binary compounds.

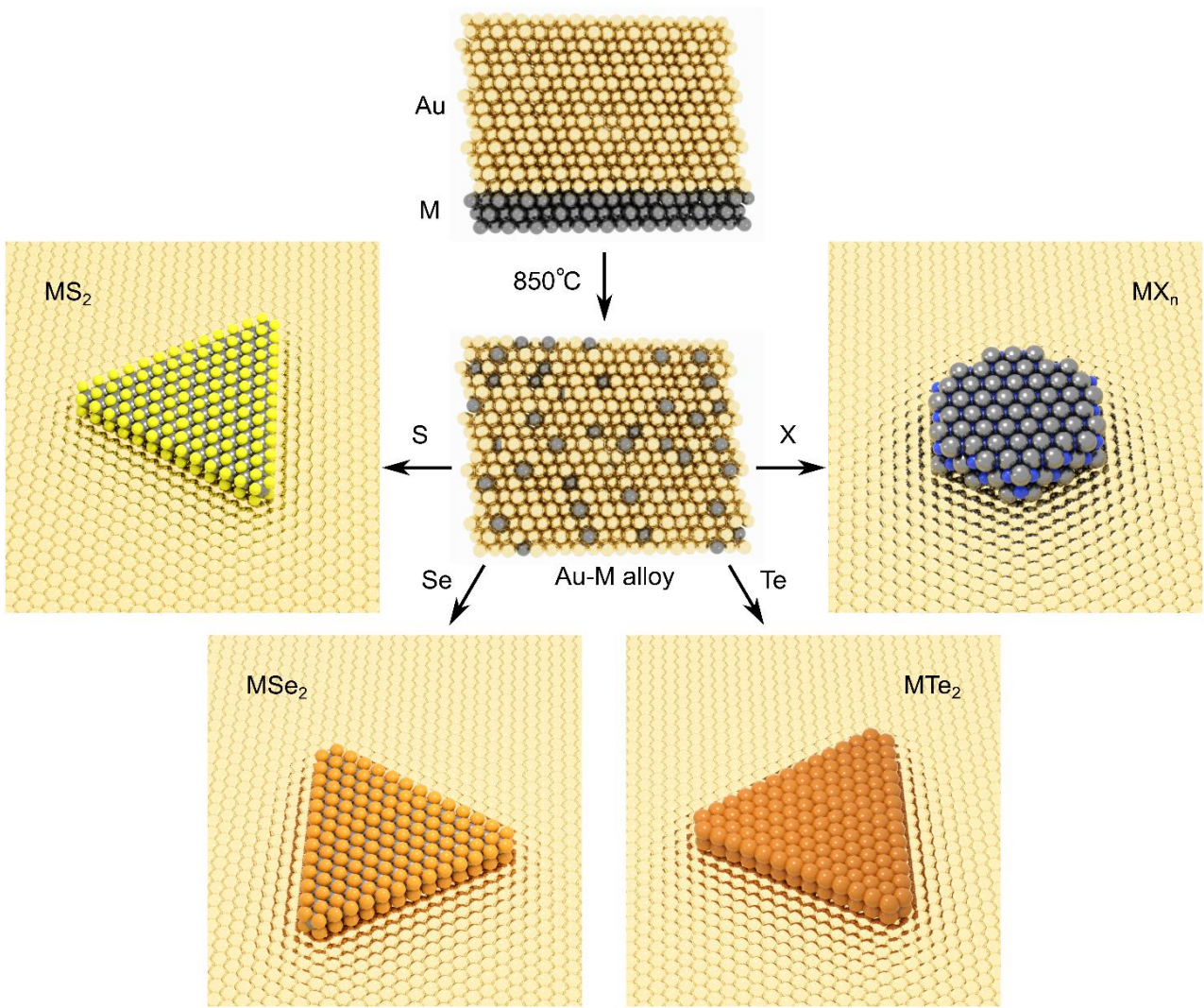

**Figure 1 – Schematic overview of the synthesis process.** A thin layer (~20 nm) of metal M is sputtered onto a *c*-plane sapphire substrate, and a thick layer (~500 nm) of Au is sputtered on top. The sample is annealed at 850 °C to produce an Au-M alloy, which is then exposed to a vapour of S, Se, Te or more generally an elemental X gas or vapour. The growth of binary $MX_n$ compounds proceeds at the surface of the Au-M layer and is surface-limited.

Of particular note is that we have used this method to synthesize a range of atomically thin binary compounds which - due to the lack of a layered bulk crystal amenable to exfoliation - have to our knowledge never been isolated or studied before.

## Results

To benchmark this approach we first synthesize and characterise $MoS_2$ layers (figure 2). Individual domains display triangular morphology as seen from scanning electron microscopy (SEM) images of the catalyst surface after growth (figure 2a). The gold catalyst adopts a {111} surface on the {001} sapphire substrate (Supplementary Fig. 1), leading to epitaxial growth of the dichalcogenide layers across the catalyst surface (figure 2b).

Raman spectroscopy for $MoS_2$ transferred to an oxidised silicon substrate with a 455 nm excitation shows two peaks, the $E_{2g}$ and $A_{1g}$ at 381.40 +/- 0.05 and 401.4 +/- 0.04 $cm^{-1}$ respectively (figure 2c). While the positions and intensities of these peaks can in general vary as a result of strain and doping, their separation of ≈ 20 $cm^{-1}$ is diagnostic of monolayer $MoS_2$[13].

High resolution and selected area electron diffraction (SAED) transmission electron microscopy (TEM) images of crystals transferred to holey carbon support grids confirms the crystal structure of the $MoS_2$ layers (figure 2d,e).

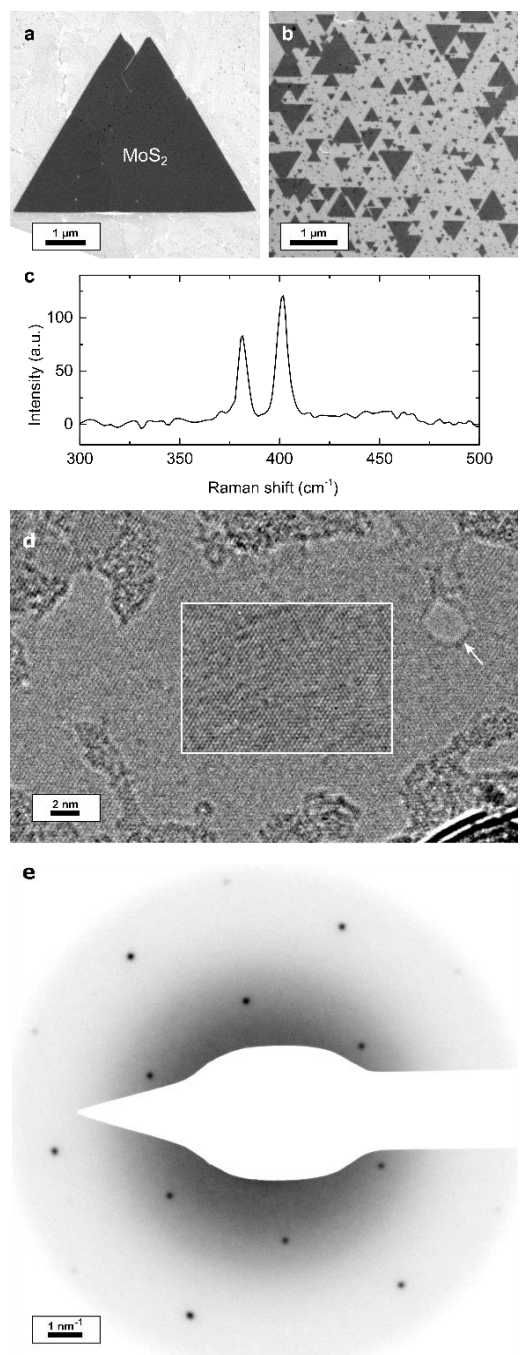

**Figure 2 – Structural characterisation of MoS$_2$. (a)** Scanning electron micrograph of an individual MoS$_2$ domain on Au {111} catalyst. **(b)** Scanning electron micrograph of epitaxially oriented MoS$_2$ domains. **(c)** Raman response with 455 nm excitation of as-grown MoS$_2$. Two peaks are evident, the E$_{2g}$ peak at 381.4 cm$^{-1}$ and the A$_{1g}$ at 401.4 cm$^{-1}$. **(d)** High resolution transmission electron micrograph of suspended MoS$_2$. A hole in the monolayer introduced by knock-on damage during imaging is indicated, where lattice

fringes are absent. The boxed region has had an iterative nonlinear denoising filter applied to highlight the MoS$_2$ lattice and reduce shot noise[14]. **(e)** Selected area diffraction patterns of suspended MoS$_2$ layers with a 100 nm diameter aperture showing single crystal long range order.

Figure 3a shows photoluminescence (PL) measurements (see Methods) of MoS$_2$ domains transferred onto oxidised silicon substrates (solid line), compared to a monolayer exfoliated MoS$_2$ crystal control sample (dashed line). The magnitude of the photoluminescence response is comparable in both cases, while the PL peak for transferred crystals is red-shifted with respect to exfoliated samples.

Electric field effect measurements were performed on unencapsulated devices of continuous MoS$_2$ films transferred onto 300 nm SiO$_2$ on Si substrates (Methods). Four-point sheet conductivity $\sigma_S$ was calculated at varying gate bias ($V_G$) as described in[15], with the results shown in figure 3b. The device shows an on-off ratio of > 10$^4$ over the gate bias range, and small maximum hysteresis of about 16V. The low level of hysteresis suggests that the overall number of charge traps is low, including intrinsic charge traps within MoS$_2$ itself[16]. The field-effect mobility µ was calculated using the standard formula[17]: $\mu = (d\sigma_S/dV_G) \cdot 1/C_{ox}$, where $C_{ox}$ is the capacitance per unit area of the back gate. Our measured devices showed a range of µ between 20 and 40 cm$^2$ V$^{-1}$s$^{-1}$, which is comparable to results for unencapsulated exfoliated MoS$_2$[17].

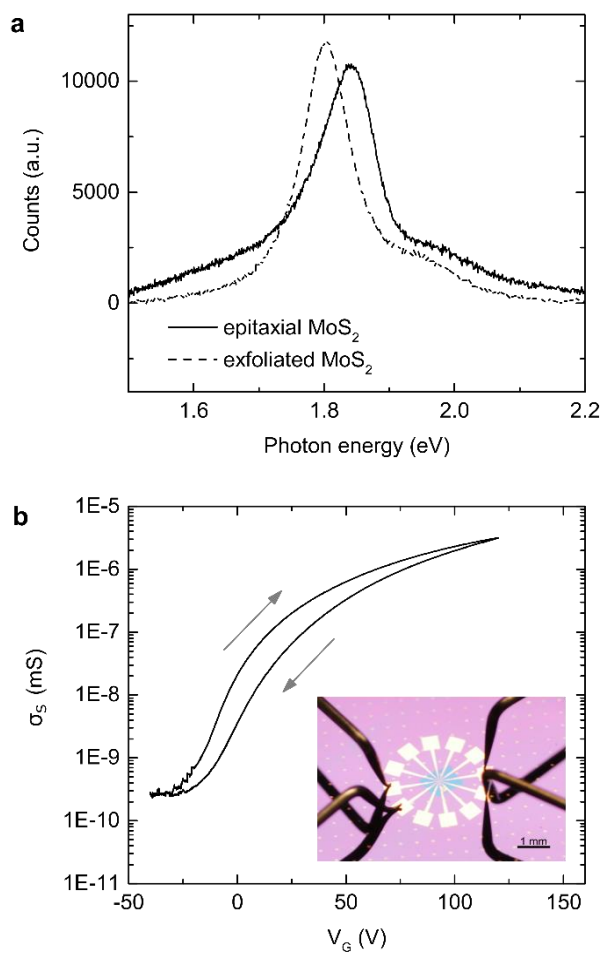

**Figure 3 – Photoluminescence and electrical characterisation of MoS$_2$. (a)** Photoluminescence results for the as-grown MoS$_2$ transferred to SiO$_2$ (solid line) vs. mechanically exfoliated monolayer MoS$_2$ (dashed line). **(b)** Electric field effect in unencapsulated MoS$_2$ devices.

Exchanging M or X for a range of different species in different combinations (M = Mo, W, Cr, Fe, Co, Hf, Nb, V; X = S, Se, Te) under identical growth conditions results in the structures visible in figure 4. XPS data confirming the expected stoichiometry and bonding for binary transition metal dichalcogenides is presented in the Supplementary Information, where further characterization on selected materials is also presented.

We have also grown few-nm thick layered structures using M = V and X = NH$_3$. XPS data show that the V:N stoichiometry is 1:1 and THz time domain spectroscopic and van der Pauw measurements of sheet conductance for these layers show a sheet resistance of 2 kΩ/□ immediately after production and transfer to SiO$_2$, rising to 10 kΩ/□ 100 days after production (see Supplementary Information).

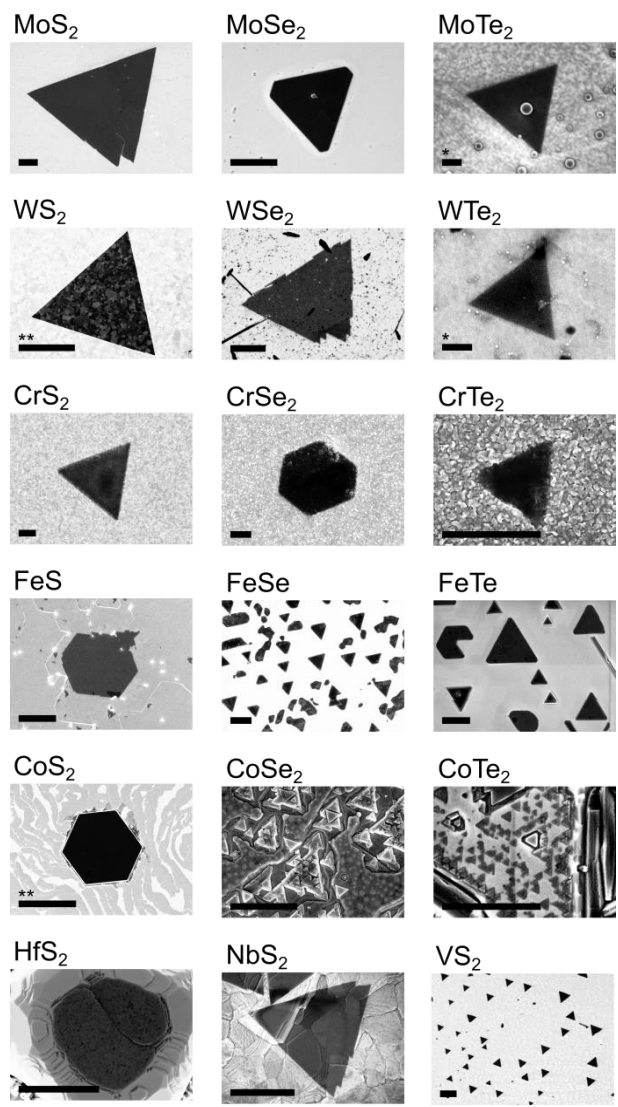

**Figure 4 – Library of 2D layered transition metal chalcogenides.** SEM images of the various transition metal chalcogenides grown by the present method. All presented

materials are grown under identical process conditions, varying only M and X. Further characterisation for the materials is presented in the Supplementary Information. Scale bars are 1 µm except where marked: *: 100 nm, **: 10 µm.

## Discussion

We have shown the scalable production of epitaxially aligned $MoS_2$ layers whose properties (namely Raman spectroscopic response, nanoscale crystalline structure, magnitude of photoluminescence and electric field effect properties) are indistinguishable from mechanically exfoliated monolayers from bulk crystals.

Furthermore, we show XPS and SEM data for a wide range of other materials produced simply by selecting M and X precursors, and additional Raman, atomic force microscopy (AFM), PL and TEM characterisation for the materials which are stable during transfer to oxidised silicon or on the catalyst layer. Without further optimisation, a large selection of materials which display morphologies, stoichiometries and chemical bonding expected of binary 2D transition metal dichalcogenides can be grown. While complete characterisation of all of these materials is beyond the scope of this manuscript - in particular for candidate materials which have not previously been synthesized or isolated from bulk crystallites before - the simplicity of this growth strategy enables very rapid experimentation in terms of the range of precursor combinations that can be tested and materials which can be grown. In practice since multiple M-Au alloy samples can be simultaneously exposed to the same X precursor, the time needed to fully characterise such materials individually far exceeds the time needed to produce them in parallel.

The surface-limited growth of a wide variety of binary 2D materials presented in this work relies on the use of gold as a catalyst layer. Gold readily alloys with most transition metals M, but shows limited solubility for X elements at our growth temperatures. Moreover, gold is unique in that it does not react with X precursors at these temperatures. Gold is catalytically active, which aids in the formation of crystalline atomically thin materials, and the {111} surface formed on *c*-plane sapphire also templates the epitaxial alignment of the grown materials. We note that similar methods for growing atomically thin carbides and sulphides have been previously reported[18,19], but only for systems where both M and X components have limited solubility (< 0.1 at.%) in the catalyst. These represent a subset of the general approach presented here, i.e. one where the solubility of M in the catalyst approaches zero. In such cases, growth is dominated by surface-mediated diffusion of both M and X. This mechanism is distinct from that presented here, where instead the growth is dominated by precipitation of component M from the catalyst bulk upon reacting with X at the solid-gas interface at the surface of the catalyst. The crucial point is that surface-limited growth does not require the metal precursor to be insoluble in the catalyst layer. This fact allows our approach to be applicable to all precursor metals, in contrast to previous reports.

In the presented work, growth targeted at complete monolayer coverage - achieved here by increasing the growth temperature - can result in adlayers. In practice adlayer growth can be reduced by tuning the growth temperature and time, reducing the flux of X, or limiting the alloy M content of the gold catalyst layer. In general we expect that optimisation of additional growth parameters for individual M-X combinations will be necessary, as has been the case for growth strategies for other 2D materials.

We highlight here that solid elemental precursors (S, Se, Te) or simple compounds (NH$_3$) are the only feedstocks required to grow the presented materials in addition to an inert gas flow. Whilst we have employed a tube furnace operating under low pressure, the growth scheme presented does not rely on a flow of X precursor (as opposed to simply the presence of X) and in principle can be performed in a sealed chamber. Such a scheme might be particularly beneficial to reduce the amount of oxygen and water in the growth system - the alloys of certain transition metals (e.g. Cr, Ta, Nb, Hf, V) are highly sensitive to oxidising impurities in the growth chamber[20,21], which can interfere with growth by passivating the catalyst surface with metal oxides. This issue can be addressed by operating under highly reducing gas flows, in a glove box, or under a suitable vacuum. Some of the materials grown also appear to be unstable under ambient conditions (i.e. V chalcogenides, Co and Fe selenides and tellurides). No special steps were taken to limit the exposure of samples to ambient atmosphere before characterisation, and even the most unstable materials presented here are stable in ambient conditions on the gold catalyst for a few hours which permits SEM and XPS characterisation. Transfer of many of these materials to e.g. oxidised silicon substrates or TEM grids is challenging however, as they rapidly degrade on contact with aqueous or oxygen-containing solutions, which hampers TEM and electrical characterisation in many cases. Recent progress in the solution-phase exfoliation of air and water-sensitive 2D crystals[22–24] suggests that similar strategies might be successfully employed here.

# Conclusion

In summary, we have demonstrated a simple and universal strategy for approaching the growth of few-atom thick binary compounds, including transition metal mono- and dichalcogenides and nitride MXenes, requiring only a volatile X precursor and grown under identical conditions. Notably, this method demonstrably enables the growth of epitaxially oriented $MX_n$ layers. New 2D materials can be made through the free choice of M and X, and different compounds can be obtained on the same growth substrate simply by switching the precursor X gas. As such, this scalable growth technique simplifies the production of existing binary 2D materials with a quality comparable to exfoliated crystals, and at the same time greatly increases the range of such materials available. We do not doubt that growth conditions for materials can be individually optimised, and that with further research, growth of in-plane and out-of-plane heterostructures could be accessible.

# Methods

**Preparation of gold-metal M substrates.** Substrates were prepared by physical vapour deposition of a thin layer (~20 nm) of metal M followed by a thick layer (between 300 nm – 1 μm, typically ~500 nm) of gold (Lesker, 99,999%) on a <001> sapphire substrate. Oxidation of the metal M is avoided by immediate encapsulation with gold before exposure to ambient conditions.

**Synthesis of 2D transition metal chalcogenides and nitrides.** Transition metal chalcogenides were synthesized in a hot-wall quartz tube reactor under low pressure conditions. The chamber was flushed three times with argon (Ar), and subsequently the samples were heated to 850°C under 100 sccm Ar. The samples were annealed at this temperature for 30-60 minutes, and growth was subsequently carried out for 10-15 minutes by exposing the samples to volatised chalcogen vapours. The vapours were generated by heating solid chalcogen precursors situated upstream from the samples: ~110°C for sulphur (sulphur flakes, Sigma Aldrich), ~220°C for selenium (selenium pellets, Sigma Aldrich), and ~420°C for tellurium (tellurium pieces, Sigma Aldrich). After growth, the samples were naturally cooled to room temperature under 100 sccm Ar flow. Synthesis of continuous MoS$_2$ films for electrical device measurements was achieved by increasing the growth temperature to 950°C.

Identical processing conditions were used to synthesise transition metal nitrides in a cold-wall reactor (AIXTRON Black Magic), except that the entire process was done under 100 sccm H$_2$ flow instead of argon in order to mitigate surface oxidation of the alloys. The growth of nitrides was performed by introducing 5 sccm NH$_3$ into the chamber for 5 minutes. The samples were then naturally cooled down to room temperature under 100 sccm H$_2$.

**Transfer of 2D materials.** Samples were transferred from the gold substrate by etching. A solution of 10% wt. PMMA in anisole (Sigma Aldrich) was spincoated onto the samples at 1500 rpm for 1 minute, after which the samples were baked at 160°C for 15 minutes. The polymer film was then manually removed at the edges of the sample. The samples were then put in a KI/I$_2$ gold etchant solution (standard gold etchant, Sigma Aldrich). After the gold was completely etched, the films were washed in DI water and transferred onto oxidised silicon substrates, where they were baked at 160°C for 10 minutes. PMMA was subsequently removed in acetone. Transfer of 2D materials onto TEM grids (Quantifoil GmbH) was done by wedging transfer[25,26] from transferred films on oxidised silicon.

**Structural characterisation.** SEM images were taken in a Zeiss Supra 40VP operated in in-lens detection mode at 5 keV. TEM characterization of transferred MoS$_2$ was done in an FEI Tecnai T20 G2 operated at 200 kV.

**Photoluminescence measurements.** Photoluminescence spectra were obtained using a custom spectroscopy setup built from a Nikon Eclipse Ti-U Inverted microscope. The excitation source was a 407 nm diode laser from Integrated Optics. The light was focused to a diffraction limited spot on the sample with a TU Plan Fluor objective from Nikon (100x, 0.9 NA) resulting in an incident power of 30 uW. The emitted fluorescent light was collected with the same objective, and the spectra were recorded using a Shamrock 303i Spectrometer equipped with a 450 nm longpass filter (FELH0450 from Thorlabs) and an electronically cooled Newton 970 EMCCD. A total of five acquisitions with 1 s exposure time each were collected for each PL spectrum.

**Fabrication of electrical devices.** MoS$_2$ regions of 1 mm$^2$ were transferred from the catalyst surface using the above procedure to silicon substrates with a 300 nm layer of thermally grown silicon dioxide and predefined electrical contacts[27]. Devices were electrically characterized in a Linkam LTS600P probe station after desorbing water from the surface[28] via baking at 225°C for 30 minutes in dry nitrogen. Subsequent measurements were performed under dry nitrogen at room temperature.

## Data Availability

The datasets generated and/or analysed during the current study are available from the corresponding author on reasonable request.

# References

1. Geim, A. K. & Grigorieva, I. V. Van der Waals heterostructures. *Nature* **499,** 419–425 (2013).
2. Anasori, B., Lukatskaya, M. R. & Gogotsi, Y. 2D metal carbides and nitrides (MXenes) for energy storage. *Nat. Rev. Mater.* **2,** 16098 (2017).
3. Tan, C. *et al.* Recent advances in ultrathin two-dimensional nanomaterials. *Chem. Rev.* **117,** 6225–6331 (2017).
4. Miró, P., Audiffred, M. & Heine, T. An atlas of two-dimensional materials. *Chem. Soc. Rev.* **43,** 6537–6554 (2014).
5. Butler, S. Z. *et al.* Progress, challenges, and opportunities in two-dimensional materials beyond graphene. *ACS Nano* **7,** 2898–2926 (2013).
6. Shi, Y., Li, H. & Li, L. Recent advances in controlled synthesis of two-dimensional transition metal dichalcogenides via vapour deposition techniques. *Chem. Soc. Rev.* **44,** 2744–2756 (2015).
7. Elías, A. L. *et al.* Controlled synthesis and transfer of large-area WS2 sheets: from single layer to few layers. *ACS Nano* **7,** 5235–5242 (2013).
8. Lee, Y.-H. *et al.* Synthesis of large-area MoS2 atomic layers with chemical vapor deposition. *Adv. Mater.* **24,** 2320–2325 (2012).
9. Lin, Y.-C. *et al.* Wafer-scale MoS2 thin layers prepared by MoO3 sulfurization. *Nanoscale* **4,** 6637–6641 (2012).
10. Wang, X. *et al.* Chemical vapor deposition growth of crystalline monolayer MoSe2. *ACS Nano* **8,** 5125–5131 (2014).
11. Kang, K. *et al.* High-mobility three-atom-thick semiconducting films with wafer-scale homogeneity. *Nature* **520,** 656–660 (2015).
12. Kidambi, P. R. *et al.* In-situ observations during chemical vapor deposition of hexagonal boron nitride on polycrystalline copper. *Chem. Mater.* **26,** 6380–6392 (2014).
13. Li, H. *et al.* From Bulk to Monolayer MoS2: Evolution of Raman Scattering. *Adv. Funct. Mater.* **22,** 1385–1390 (2012).
14. Du, H. A nonlinear filtering algorithm for denoising HR(S)TEM micrographs. *Ultramicroscopy* **151,** 62–67 (2015).
15. Mackenzie, D. M. A. *et al.* Quality assessment of graphene: Continuity, uniformity, and accuracy of mobility measurements. *Nano Res.* **10,** 3596–3605 (2017).
16. Kaushik, N. *et al.* Reversible hysteresis inversion in MoS2 field effect transistors. *npj 2D Mater. Appl.* **1,** 34 (2017).
17. Kim, J. H. *et al.* Thickness-dependent electron mobility of single and few-layer MoS 2 thin-film transistors. *AIP Adv.* **6,** 65106 (2016).
18. Xu, C. *et al.* Large-area high-quality 2D ultrathin Mo2C superconducting crystals. *Nat. Mater.* **14,** 1135–1141 (2015).
19. Zhang, T. *et al.* Twinned growth behaviour of two-dimensional materials. *Nat. Commun.* **7,** 13911 (2016).
20. Tynkova, A., Sidorenko, S., Voloshko, S., Rennie, A. R. & Vasylyev, M. A. Interdiffusion in Au (120 nm)/Ni (70 nm) thin films at the low-temperature annealing in the different atmospheres. *Vacuum* **87,** 69–74 (2013).
21. Ashwell, G. W. B. Interdiffusion of Titanium and Gold: A Comparison of Thin Films Deposited in Technical Vacuum and Ultrahigh Vacuum. *J. Electrochem. Soc.* **128,**

649–654 (1981).
22. Hanlon, D. *et al.* Liquid exfoliation of solvent-stabilized few-layer black phosphorus for applications beyond electronics. *Nat. Commun.* **6,** 8563 (2015).
23. Kang, J. *et al.* Solvent exfoliation of electronic-grade, two-dimensional black phosphorus. *ACS Nano* **9,** 3596–3604 (2015).
24. Kang, J. *et al.* Stable aqueous dispersions of optically and electronically active phosphorene. *Proc. Natl. Acad. Sci.* **113,** 11688–11693 (2016).
25. Schneider, G. F., Calado, V. E., Zandbergen, H., Vandersypen, L. M. K. & Dekker, C. Wedging transfer of nanostructures. *Nano Lett.* **10,** 1912–1916 (2010).
26. Thomsen, J. D. *et al.* Suppression of intrinsic roughness in encapsulated graphene. *Phys. Rev. B* **96,** 14101 (2017).
27. Mackenzie, D. M. A. *et al.* Graphene antidot lattice transport measurements. *Int. J. Nanotechnol.* **14,** 226 (2017).
28. Gammelgaard, L. *et al.* Graphene transport properties upon exposure to PMMA processing and heat treatments. *2D Mater.* **1,** 35005 (2014).

## Acknowledgements

A.S., J.D.T., T.B. and P.B. acknowledge support from the EU Seventh Framework Programme (FP7/2007-2013) under grant agreement number FP7-6040007 "GLADIATOR". P.R.W and P.B. acknowledge the support from the EC Graphene FET Flagship, grant agreement number 604391. M.G., T.B., and P.B. acknowledge support from the Danish National Research Foundation Center of Excellence for Nanostructured Graphene (CNG), project DNRF103. A.C. acknowledges support from the Villum Fonden, research grant (9455).

## Author Contributions

A.S. conceived the method and performed the majority of the experiments. J.D.T. and T.B. performed TEM and SAED measurements. J.K. performed STEM characterisation on VN lamellas, while Z.I.B. prepared the FIB lamellas. A.C. did ellipsometry measurements. P.R.W. performed terahertz time domain spectroscopy measurements. M.G. did photoluminescence measurements. T.B. and P.B. supervised the project. T.B. and A.S. wrote the manuscript and prepared the figures, and all authors contributed to manuscript revisions.

## Corresponding Author

*(T.B.) e-mail: tim.booth@nanotech.dtu.dk

## Notes

The authors declare no competing financial interest.

Supplementary Information for

# A general approach for the synthesis of two-dimensional binary compounds

Abhay Shivayogimath[1,2], Joachim Dahl Thomsen[1,2], David M. A. Mackenzie[1,2], Mathias Geisler[2,3], Jens Kling[4], Zoltan Imre Balogh[4], Andrea Crovetto[1,5], Patrick R. Whelan[1,2], Peter Bøggild[1,2], Timothy J. Booth[1,2]*

[1]*DTU Nanotech, Technical University of Denmark, Ørsteds Plads 345E, DK-2800 Kgs. Lyngby, Denmark*

[2]*Centre for Nanostructured Graphene (CNG), Technical University of Denmark, Ørsteds Plads 345C, DK-2800 Kgs. Lyngby, Denmark*

[3]*DTU Fotonik, Technical University of Denmark, Ørsteds Plads 343, DK-2800 Kgs. Lyngby, Denmark*

[4]*DTU Cen (Center for Electron Nanoscopy), Technical University of Denmark, Fysikvej 307, DK-2800 Kgs. Lyngby, Denmark*

[5]*V-SUSTAIN, Villum Center for the Science of Sustainable Fuels and Chemicals, Technical University of Denmark, DK-2800 Kgs. Lyngby, Denmark*

*email: tim.booth@nanotech.dtu.dk

## Supplementary Methods

**Sample characterisation.** Optical microscopy images were taken in a Nikon Eclipse L200N optical microscope. Scanning electron microscopy (SEM) was conducted in a Zeiss Supra 40VP operated in in-lens detection mode at 5.00 keV. Atomic force microscopy (AFM) scans were done in a Bruker AFM Dimension Icon. X-ray photoelectron spectroscopy (XPS) measurements were conducted in a Thermo Fisher Scientific K-alpha XPS system using an Al Kα X-ray source (1486.7eV). Acquisition parameters were kept constant for all survey and high-resolution scans (300 ms x 10 scans). Raman spectroscopy was conducted in a Thermo Fisher DXR microscope equipped with a 455 nm laser (5 mW, 10 s x 5 exposure time, 50x objective).

**Electron backscatter diffraction (EBSD).** EBSD measurements were used to determine the surface grain orientation of Au-M alloys after processing, which in the case of Supplementary Figure 1 was an Au-V alloy. Measurements were done in a FEI Nova NANO SEM 600 using the following parameters: 15 kV, 12 nA, 10.5 mm working distance, 50 μm aperture, a 10 μm x 10 μm scan area and a step size of 280 nm. The obtained diffraction patterns were indexed against pure Au and Au-V (85%-15%) structures, and were found to match with the pure Au phase only. Various different regions were also scanned to ensure homogeneity throughout the sample. The SEM image in Supplementary Figure 1 was taken in the same instrument, using a secondary electron detector.

**Terahertz-time domain spectroscopy (THz-TDS).** THz-TDS mapping of the sheet conductivity of VN on quartz was acquired using a Picometrix T-ray 4000 fibre-coupled spectrometer. The sample was raster-scanned to form a spatial map with 200 μm step size and a resolution of ≈ 350 μm at 1 THz. The frequency-dependent sheet conductivity was determined as described in detail in previous reports[1,2].

**Ellipsometry.** Ellipsometry measurements were performed in reflection mode with a rotating compensator spectroscopic ellipsometer (M-2000, J.A. Woollam Co.). The measurement of the ellipsometric quantities ψ and Δ in the spectral range 0.7-5.9 eV was repeated at seven angles of incidence (45-75°) at each measurement spot using a collimated beam with a spot size of approximately 200 x 300 μm. Four spots were measured on the transferred VN film on quartz. The multi-angle ψ and Δ spectra for each spot were fitted simultaneously to an optical model, with thickness and optical functions of VN being the unknown parameters. The optical functions of the quartz substrate were determined in a separate ellipsometry measurement. A multi-sample analysis routine was employed to accurately determine both the thickness and the optical functions of VN avoiding the risk of cross-correlation between fitted parameters. In one of the four spots, the thickness was fixed to the value of 5.5 nm as determined by both AFM and cross-sectional TEM. The thicknesses of the three remaining spots were fitted independently.

The optical functions of VN were parameterised as explained in the caption of Supplementary Figure 24, but were not allowed to vary from spot to spot.

**Transmission electron microscopy.** TEM characterisation of W, Mo chalcogenides and $CoS_2$ was done in an FEI Tecnai T20 G2 operated at 200 kV. Bright field HRTEM images of transferred VN films were taken in an FEI Titan 80-300 environmental TEM with post-specimen spherical aberration correction, operated at 80 kV. Scanning transmission electron microscopy (STEM) images were acquired in an FEI Titan 80-300 with probe corrector at 300kV. A standard dark-field detector on the Gatan GIF entrance aperture with a short camera length of indicated 38 mm and a half-convergence angle of about 17.6 degree were used. Cross-sectional lamella of as-grown VN on gold was made by focussed ion beam milling in an FEI Helios EBS3 microscope. The lamella thickness prior to STEM imaging was 100-150 nm, as measured by EELS.

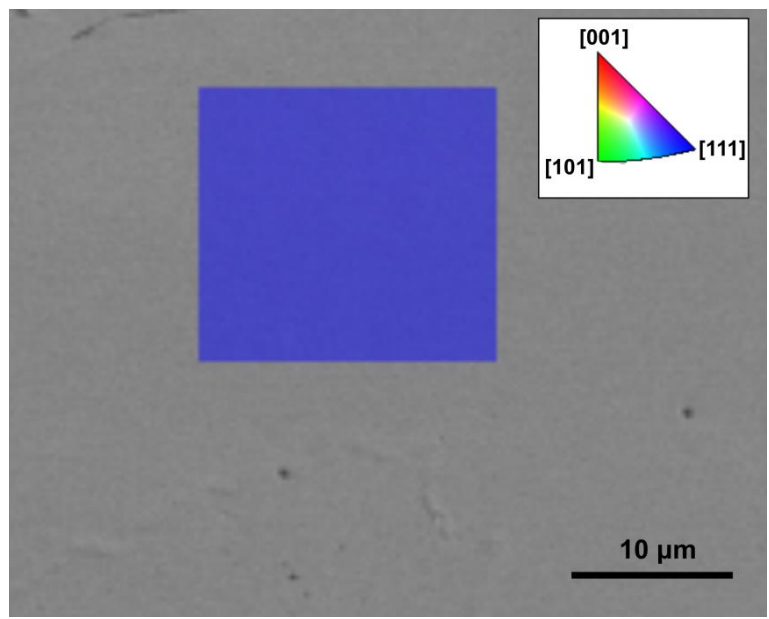

**Supplementary Figure 1. EBSD of post-process Au alloys on c-plane sapphire.** EBSD map of a ~100 µm² area overlaid onto an SEM image of the alloy surface. The EBSD clearly shows that the surface has the Au {111} orientation.

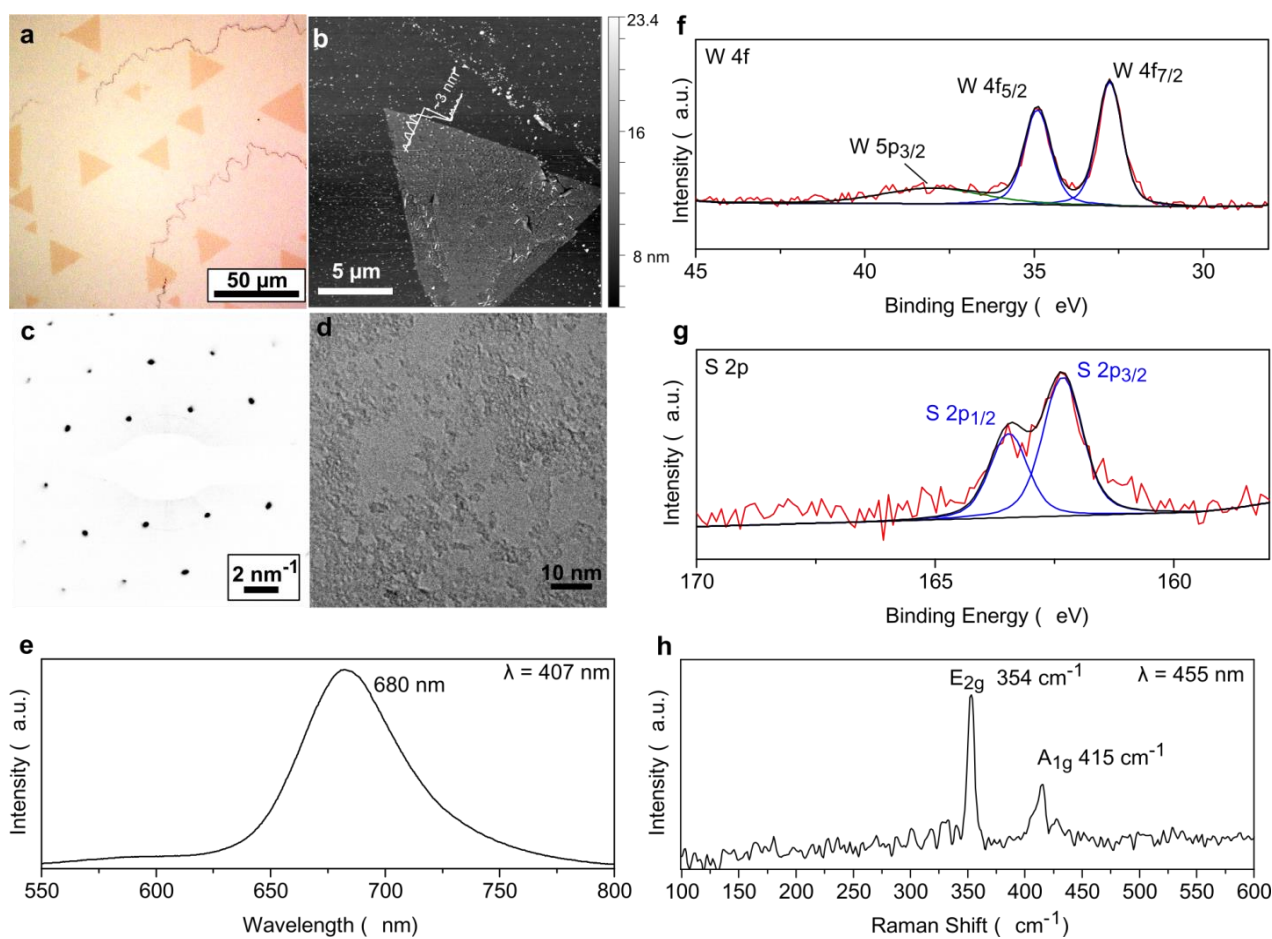

**Supplementary Figure 2. Characterisation of tungsten disulphide (WS$_2$). (a)** Optical image of WS$_2$ domains on gold, showing evidence of epitaxial alignment with the underlying substrate. **(b)** Atomic force microscopy (AFM) image of a WS$_2$ domain transferred onto 90 nm SiO$_2$ on Si substrate; the measured thickness of the domain is ~ 3 nm, which is within the thickness of a monolayer when taking into account polymer residues, surface oxidation, and tip and substrate interactions[3,4]. **(c)** Selected area diffraction (SAED) pattern and **(d)** bright field TEM of a WS$_2$ domain transferred onto TEM grids. **(e)** Photoluminescence spectrum of WS$_2$ domains transferred onto 90 nm SiO$_2$/Si substrates, showing a strong peak at 680 nm. The peak is redshifted from reported values for monolayer WS$_2$[5], which we suspect is due to screening and doping effects from transfer residues. **(f-g)** High resolution XPS spectra of **(f)** W 4f and **(g)** S 2p regions of as-grown WS$_2$ on the gold surface, where the peaks at 32.8 eV (W 4f$_{7/2}$), 34.9 eV (W 4f$_{5/2}$), 162.3 eV (S 2p$_{3/2}$) and 163.5 eV (S 2p$_{1/2}$) correspond to WS$_2$ (NIST XPS database). **(h)** Raman spectra of WS$_2$ domains on gold, indicating monolayer thickness[6].

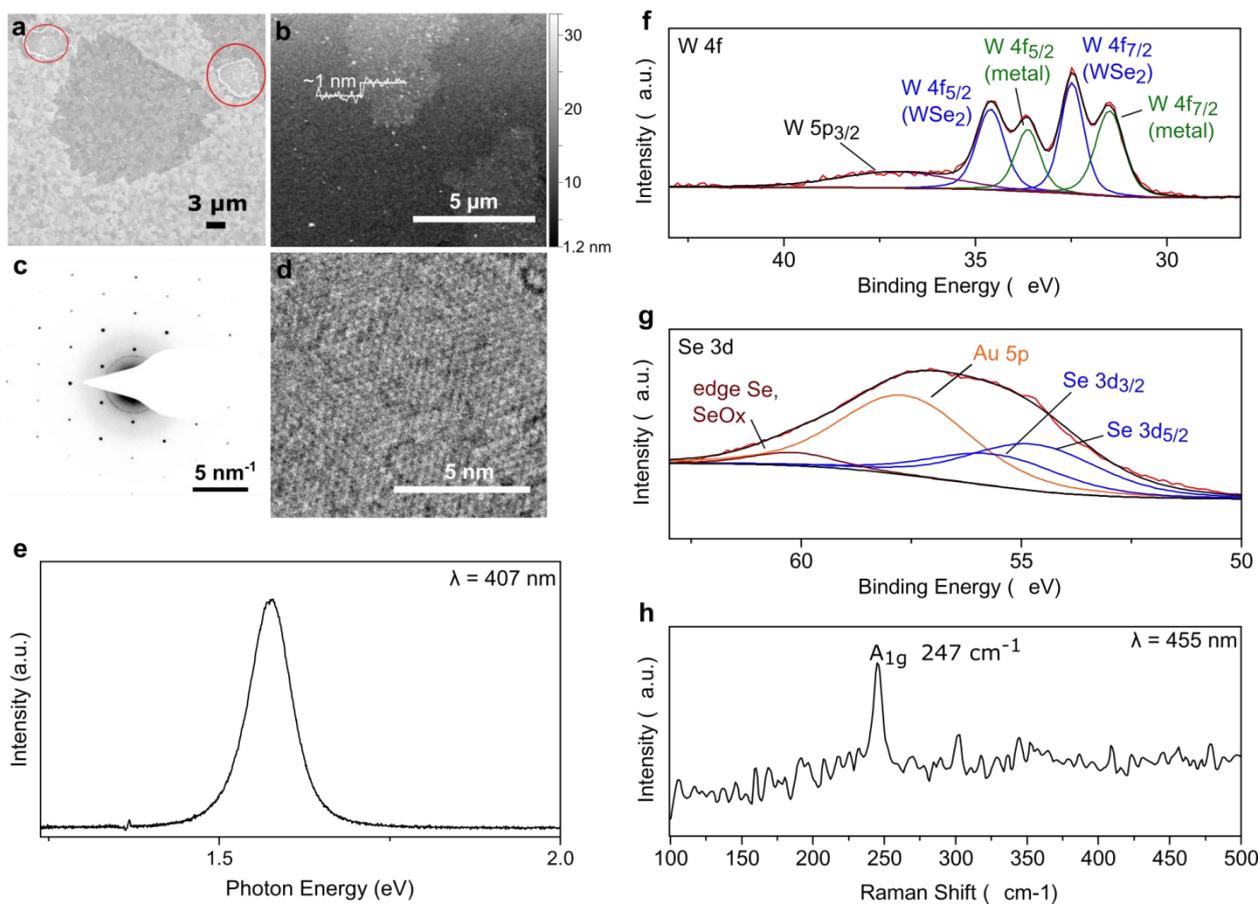

**Supplementary Figure 3. Characterisation of tungsten diselenide (WSe$_2$). (a)** SEM image of WSe$_2$ domains on gold. Red circles in **(a)** indicate areas where the gold film has dewetted the underlying tungsten layer due to its thickness (initial thickness of the gold film here was 300 nm). **(b)** AFM image of a WSe$_2$ domain transferred onto 90 nm SiO$_2$/Si substrate; the measured thickness of the domain is ~ 1 nm, not accounting for transfer residues and chemical contrast between tip and substrate. **(c)** SAED pattern and **(d)** bright field TEM of WSe$_2$ transferred onto TEM grids. **(e)** Photoluminescence spectrum of WSe$_2$ domains transferred onto 90 nm SiO$_2$/Si substrates, which shows a strong peak at 1.57 eV. **(f-g)** High resolution XPS spectra of **(f)** W 4f and **(g)** Se 3d regions of as-grown WSe$_2$ on the gold surface. Peaks at 32.5 eV and 34.6 eV in (f) correspond to WSe$_2$, while the peaks at 31.5 eV and 33.6 eV correspond to the exposed W metal in the dewetted areas in (a) (NIST XPS Database). Peaks in (g) at 54.7 eV and 55.6 eV correspond to Se 3d peaks of WSe$_2$, while we attribute the peak at 60.2 eV to oxidised/edge selenides[7,8]. **(h)** Raman spectra of WSe$_2$ domains on gold, which indicates that they are monolayers[9].

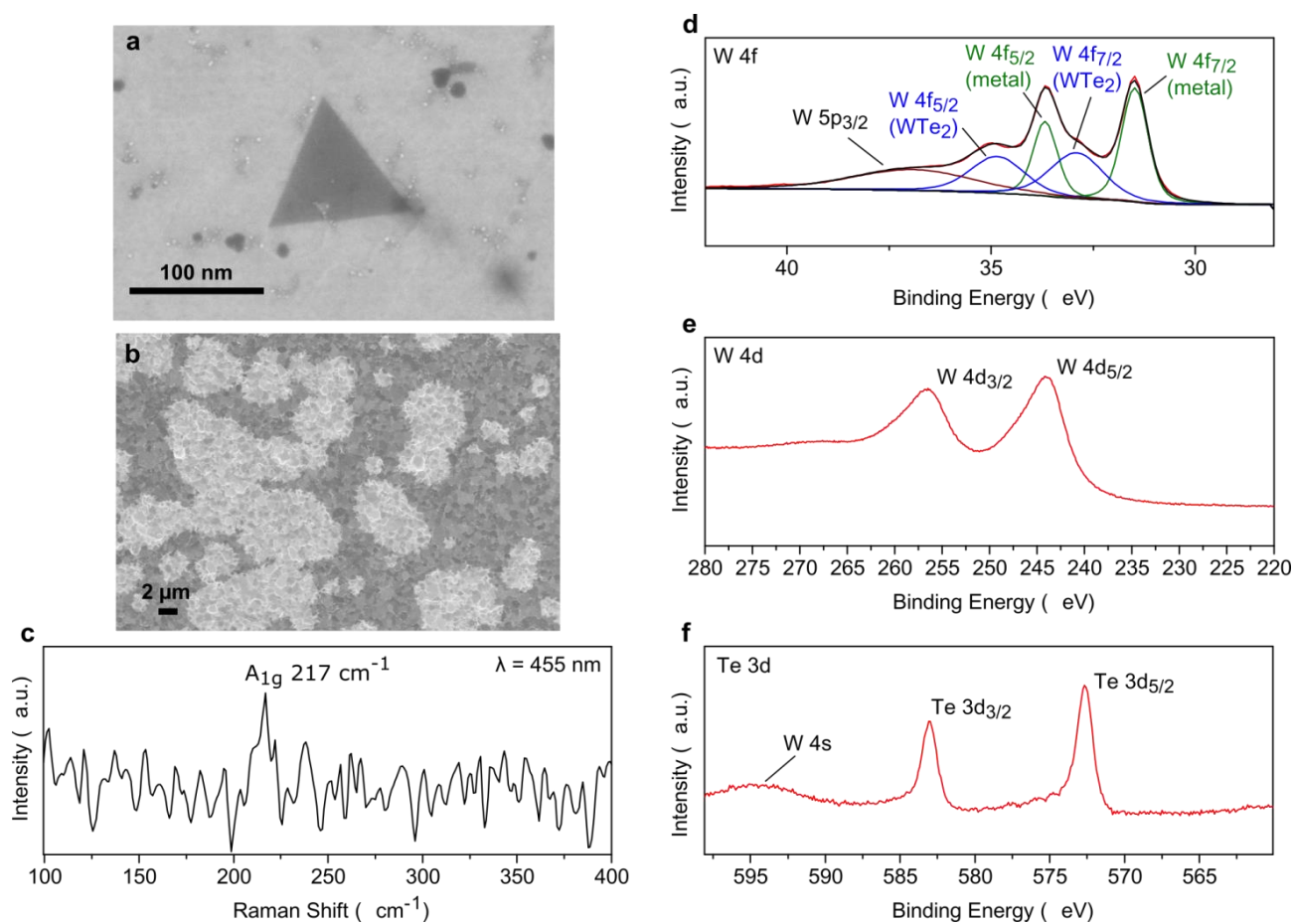

**Supplementary Figure 4. Characterisation of tungsten ditelluride (WTe$_2$). (a)** High magnification of isolated WTe$_2$ domains and **(b)** low magnification SEM images of merged WTe$_2$ domains on gold. Dark coloured regions in **(b)** are partially grown WTe$_2$ films. **(c)** Raman spectra of WTe$_2$ on gold, which indicates that the film is a monolayer[10], **(d-f)** High resolution XPS spectra of **(d)** W 4f, **(e)** W 4d and **(f)** Te 3d regions of as-grown WTe$_2$ on gold. W 4f peaks in (d) at 32.9 eV and 34.85 eV correspond to WTe$_2$, whereas the peaks at 31.5 eV and 33.7 eV correspond to W metal from dewetted areas, similar to the case of WSe$_2$ (the initial gold film thickness was 300 nm in this case as well). W 4d peaks in (e) at 244.2 eV and 256.5 eV correspond to reported values for WTe$_2$[11]. Te 3d peaks in (f) are at 572.7 eV and 583 eV, corresponding to WTe$_2$[11]. WTe$_2$ films did not appear to survive transfer onto 90 nm SiO$_2$ on Si substrates.

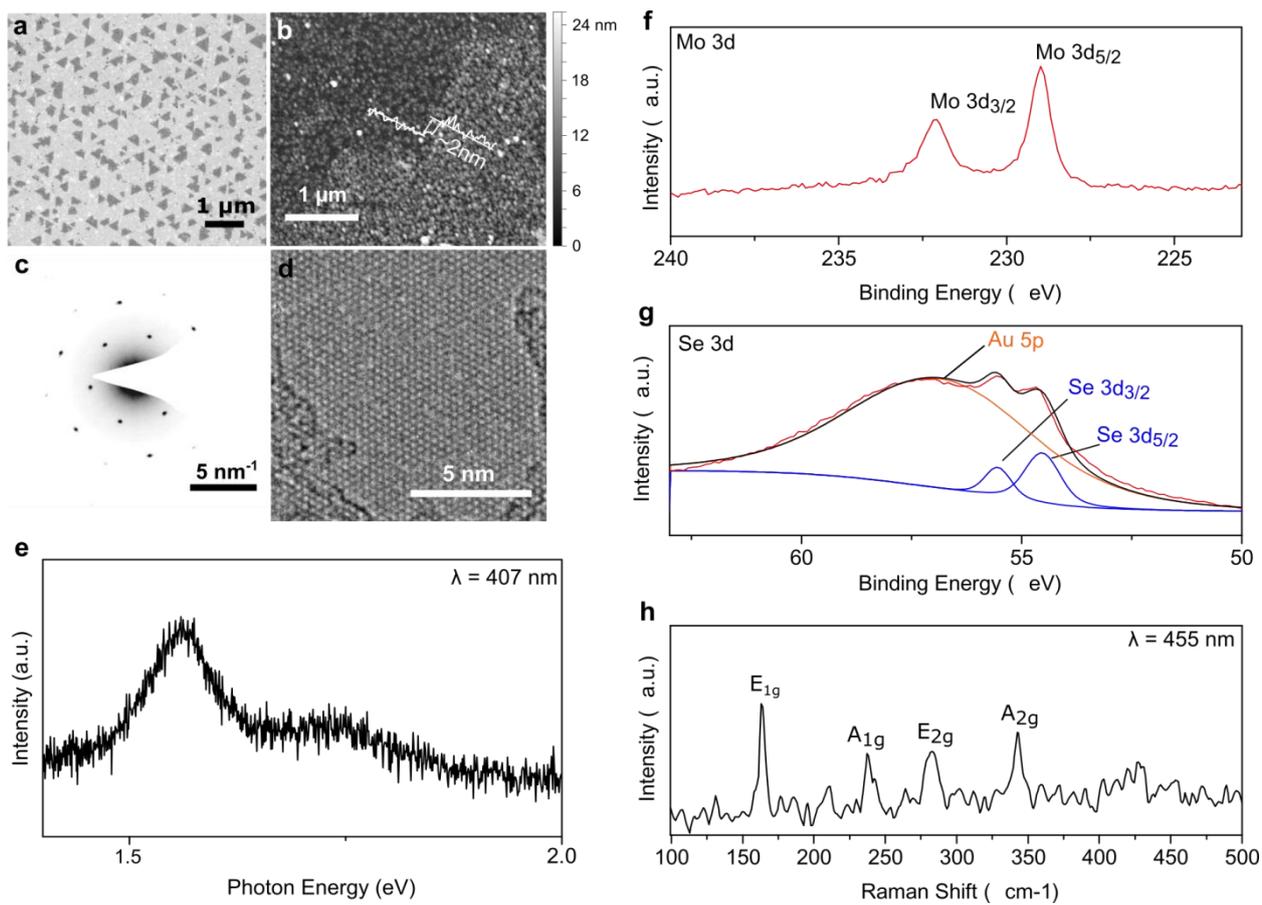

**Supplementary Figure 5. Characterisation of molybdenum diselenide (MoSe$_2$). (a)** SEM image of MoSe$_2$ domains on gold, showing evidence of epitaxial alignment with the underlying gold. **(b)** AFM on MoSe$_2$ domains transferred onto 90 nm SiO$_2$/Si substrate; the measured thickness of the domain is ~ 2 nm, not accounting for transfer residues and chemical contrast between tip and substrate. **(c)** SAED pattern and **(d)** high resolution TEM image of MoSe$_2$ domains transferred onto TEM grids. Visible defects are caused by knock-on damage during imaging at 200 kV. **(e)** Photoluminescence spectrum of MoSe$_2$ domains transferred onto 90 nm SiO$_2$/Si substrates, showing A and B exciton peaks at 1.57 eV and 1.75 eV, respectively[12]. **(f-g)** High resolution XPS spectra of the **(f)** Mo 3d and **(g)** S 2p regions of MoSe$_2$ on gold. Mo 3d peaks in (f) at 228.9 eV and 232.1 eV and Se 3d peaks in (g) at 54.6 eV and 55.5 eV correspond to MoSe$_2$[13]. **(h)** Raman spectra of MoSe$_2$ transferred onto 90 nm SiO$_2$/Si substrate; the pronounced E$_{1g}$, E$_{2g}$, and A$_{2g}$ peaks are attributed to the excitation wavelength used in this work[14].

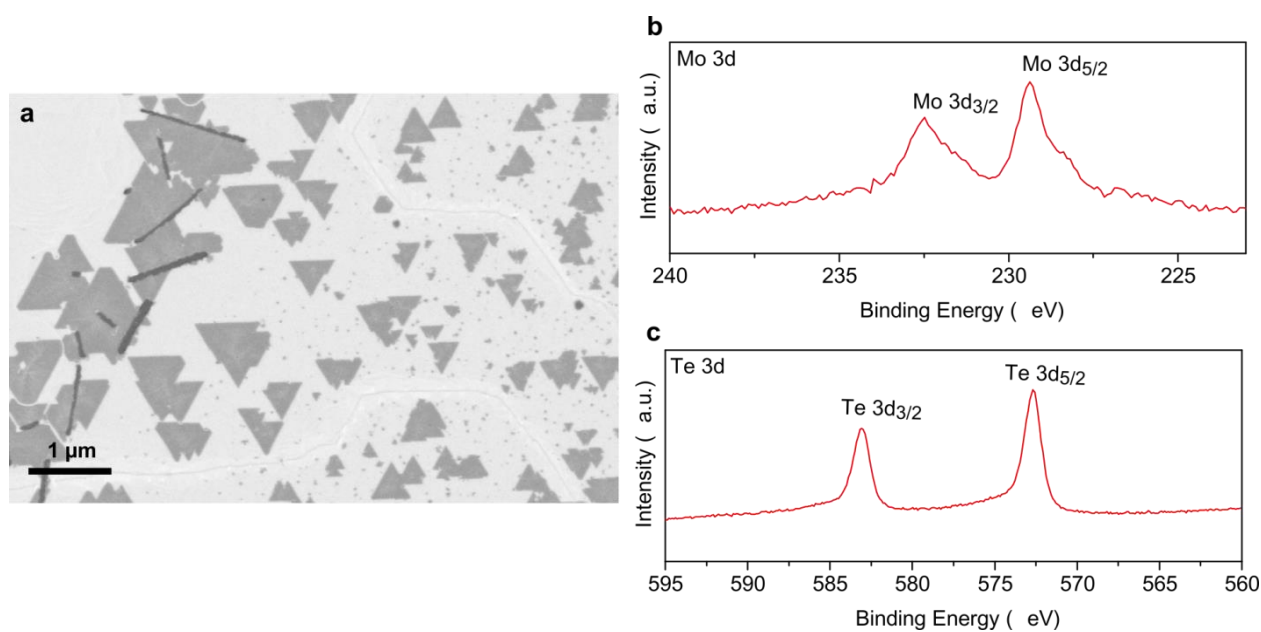

**Supplementary Figure 6. Characterisation of molybdenum ditelluride (MoTe$_2$). (a)** SEM image of a MoTe$_2$ domain on gold. **(b-c)** High resolution XPS spectra of the **(b)** Mo 3d and **(c)** Te 3d regions of MoTe$_2$ on gold. Mo 3d peaks in (b) at 229.2 eV and 232.3 eV and Te 3d peaks in (c) at 572.6 eV and 582.9 eV correspond to MoTe$_2$[15]. We were unable to detect a sufficiently strong Raman signal from the domains on gold with our experimental setup, and the domains did not appear to survive transfer onto SiO$_2$ substrates.

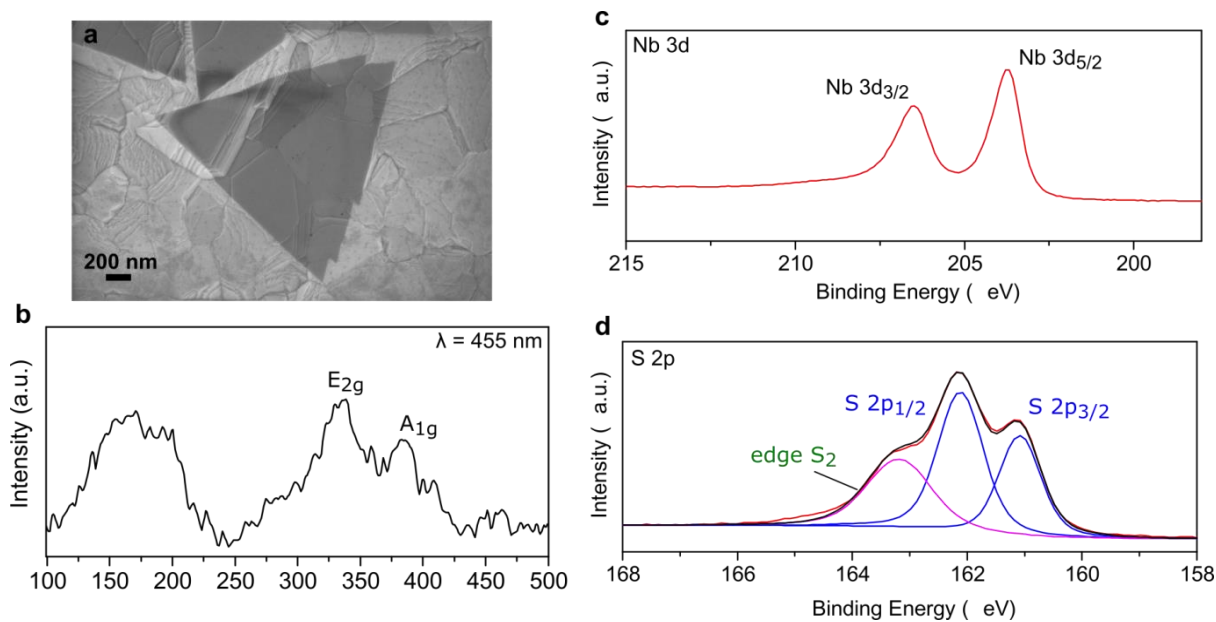

**Supplementary Figure 7. Characterisation of niobium disulphide (NbS$_2$). (a)** SEM image of an NbS$_2$ domain on gold. (b) Raman spectrum of an NbS$_2$ film on gold, where the E$_{2g}$ and A$_{1g}$ correspond to reported values for NbS$_2$[16]. High resolution XPS spectra of the **(c)** Nb 3d and **(d)** S 2p regions of NbS$_2$ on gold. Nb 3d peaks in (c) are at 203.7 eV and 206.5 eV and S 2p peaks in (d) are at 161.1 eV, 162.2 eV and 163.2 eV.

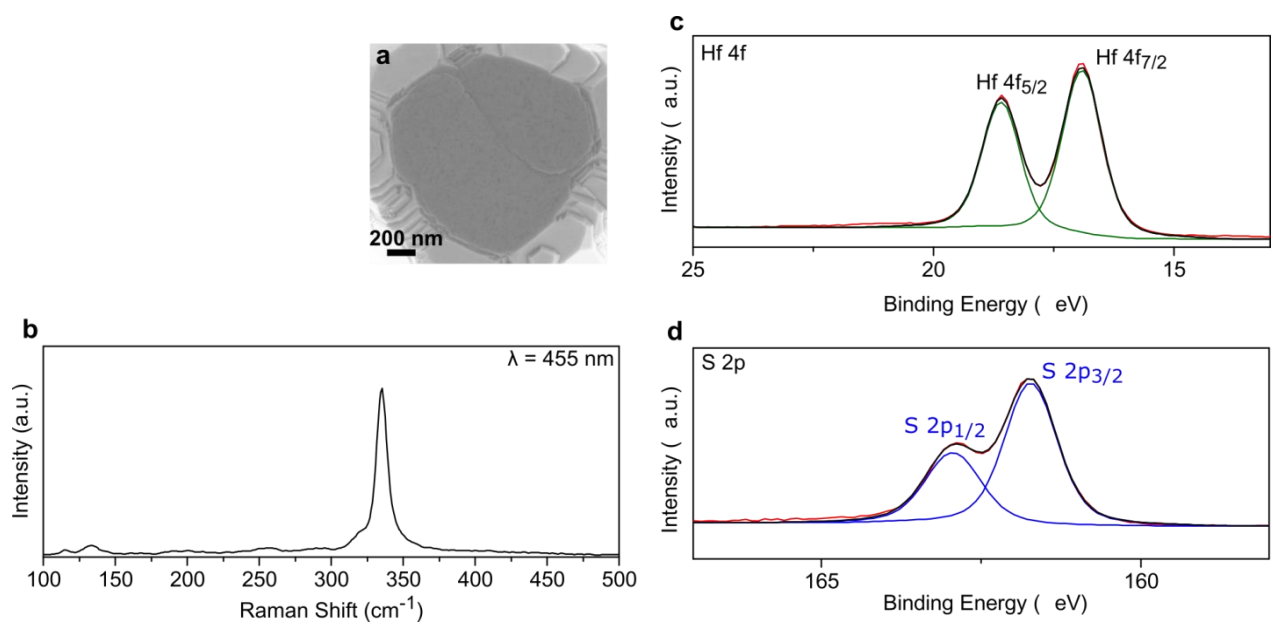

**Supplementary Figure 8. Characterisation of hafnium disulphide (HfS$_2$). (a)** High magnification and of an isolated HfS$_2$ domain on gold. **(b)** Raman spectrum of an HfS$_2$ film on gold, which shows a pronounced peak at ~335 cm$^{-1}$ corresponding to reported values for HfS$_2$[17]. **(c-d)** High resolution XPS spectra of the **(c)** Hf 4f and **(d)** S 2p regions of HfS$_2$ on gold; Hf 4f peaks are at 16.9 eV and 18.6 eV and S 2p peaks are at 161.7 eV and 162.9 eV. The Hf:S ratio here is 1:2.

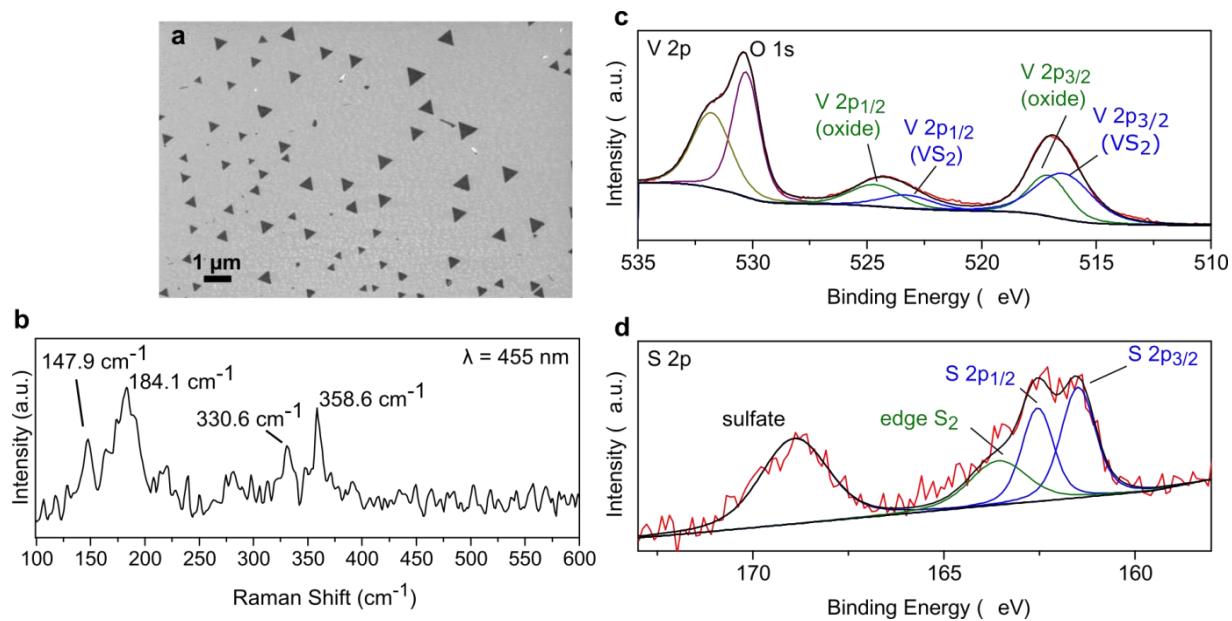

**Supplementary Figure 9. Characterisation of vanadium disulphide (VS$_2$). (a)** SEM images of VS$_2$ domains on gold, showing evidence of epitaxial alignment with the underlying Au (111) substrate. **(b)** Raman spectrum of VS$_2$ film on gold, where the peaks correspond to reported values for VS$_2$ nanosheets[18]. **(c-d)** XPS spectra of the gold surface post-growth: **(c)** V 2p and **(d)** S 2p scan. The V 2p spectrum shows peaks at 516.4 eV and 523.3 eV, which correspond to the V$^{4+}$ oxidation state in VS$_2$[19]. The peaks at 517.1 eV and 524.7 eV can be attributed to vanadium oxides[7]. The S 2p peaks at 161.5 eV and 162.6 eV are attributed to VS$_2$. VS$_2$ is air-sensitive and degrades upon exposure to ambient, as evidenced by the prominent sulphate peak in (d).

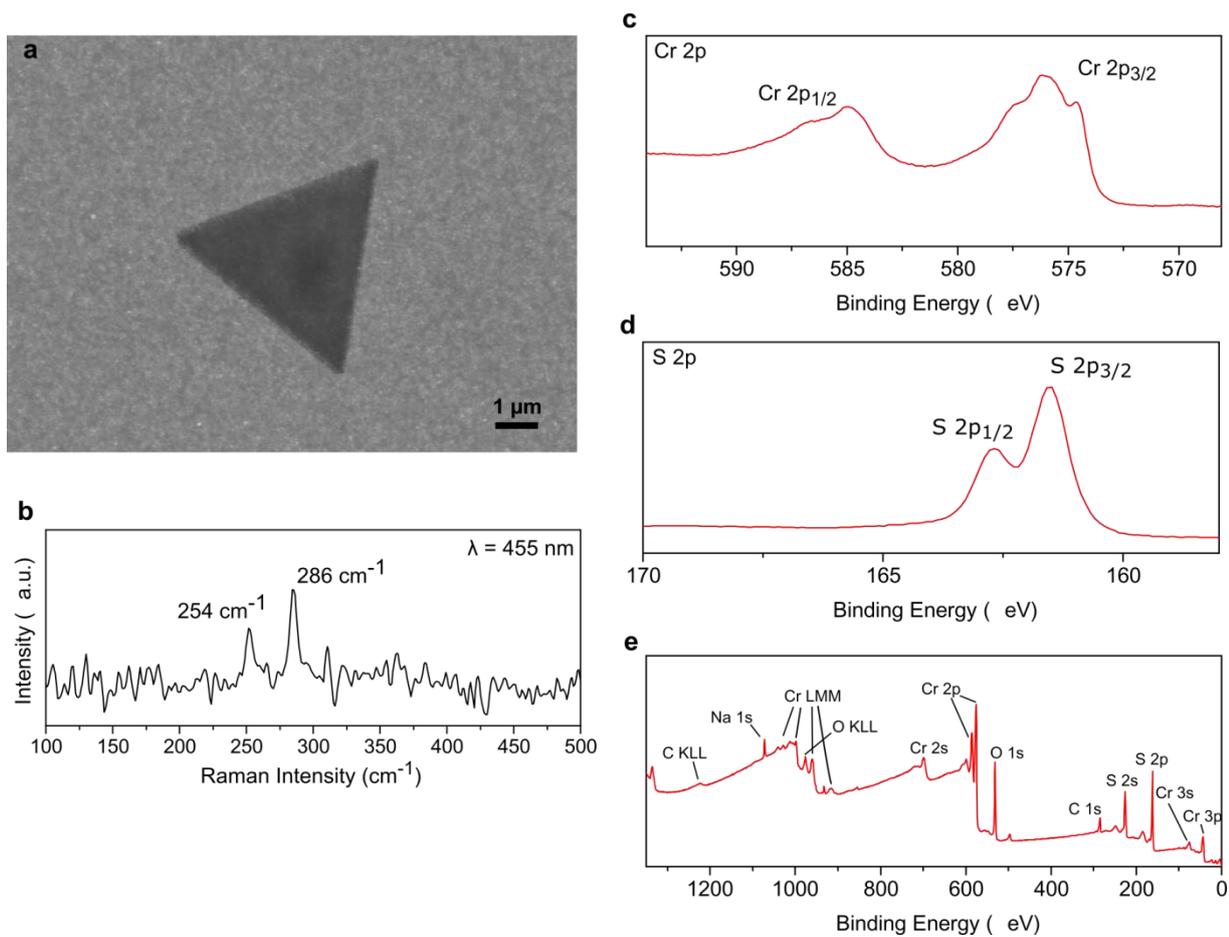

**Supplementary Figure 10. Characterisation of chromium disulphide (CrS$_2$). (a)** SEM image of a CrS$_2$ domain on gold. **(b)** Raman spectrum of a CrS$_2$ domain on gold. **(c-e)** XPS spectra of the gold surface post-growth: **(c)** Cr 2p, **(d)** S 2p, and **(e)** survey scan.

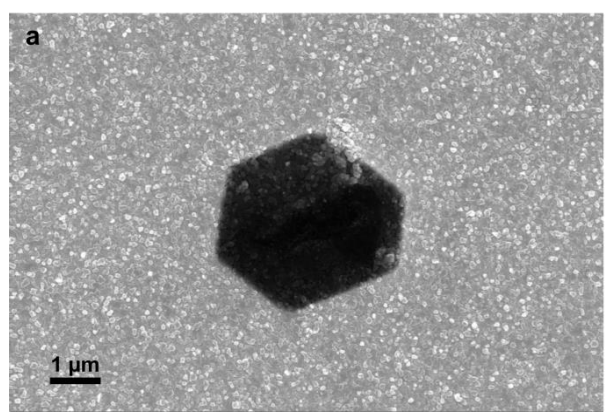
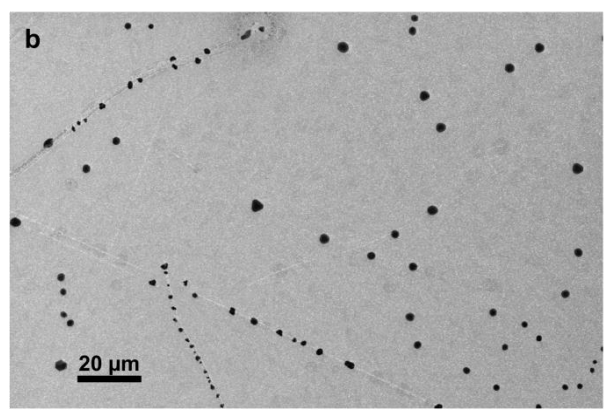
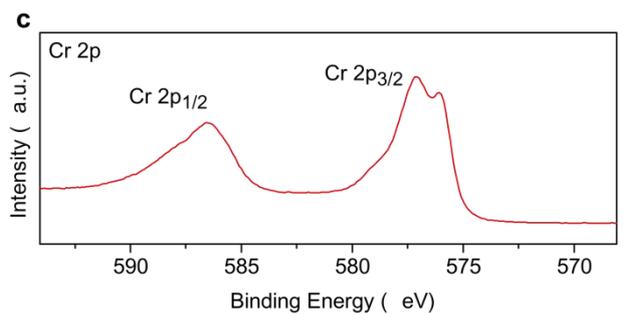
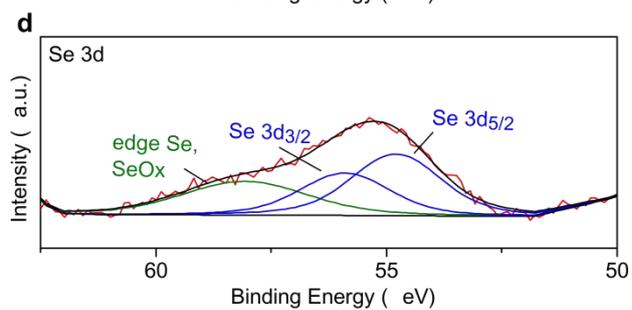
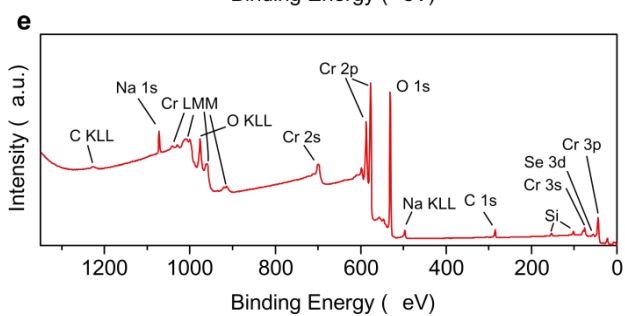

**Supplementary Figure 11. Characterisation of chromium diselenide ($CrSe_2$). (a)** High magnification and **(b)** low magnification SEM images of $CrSe_2$ on gold. **(c-e)** XPS spectra of the gold surface post-growth: **(c)** Cr 2p, **(d)** Se 3d, and **(e)** survey scan.

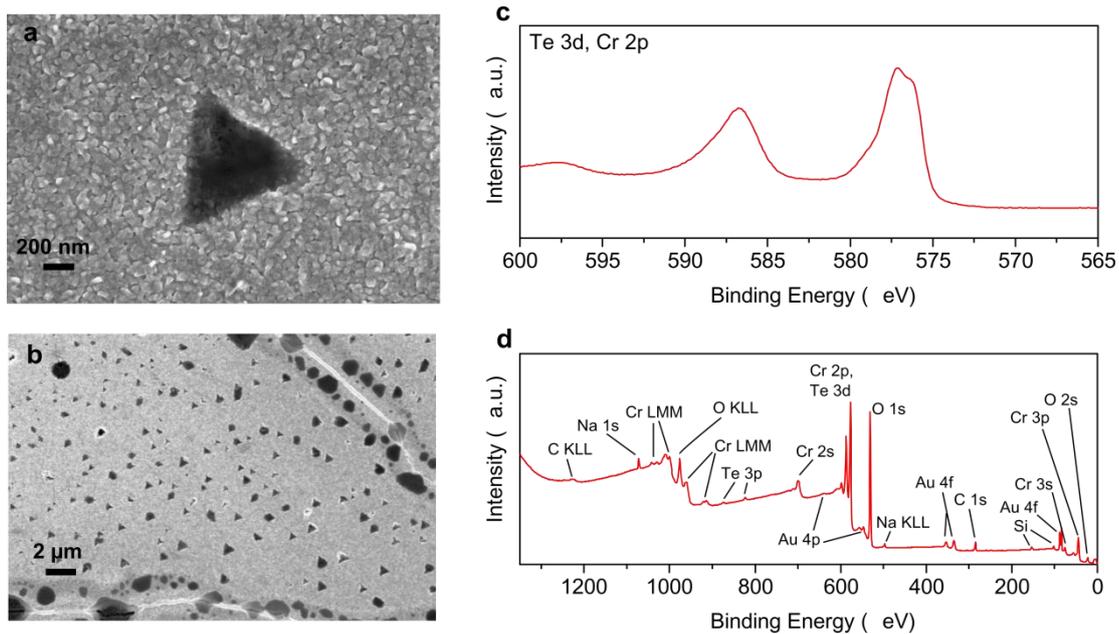

**Supplementary Figure 12. Characterisation of chromium ditelluride (CrTe$_2$). (a)** High magnification and **(b)** low magnification SEM images of CrTe$_2$ on gold. **(c-d)** XPS spectra of the gold surface post-growth: **(c)** Te 3d, Cr 2p and **(e)** survey scan.

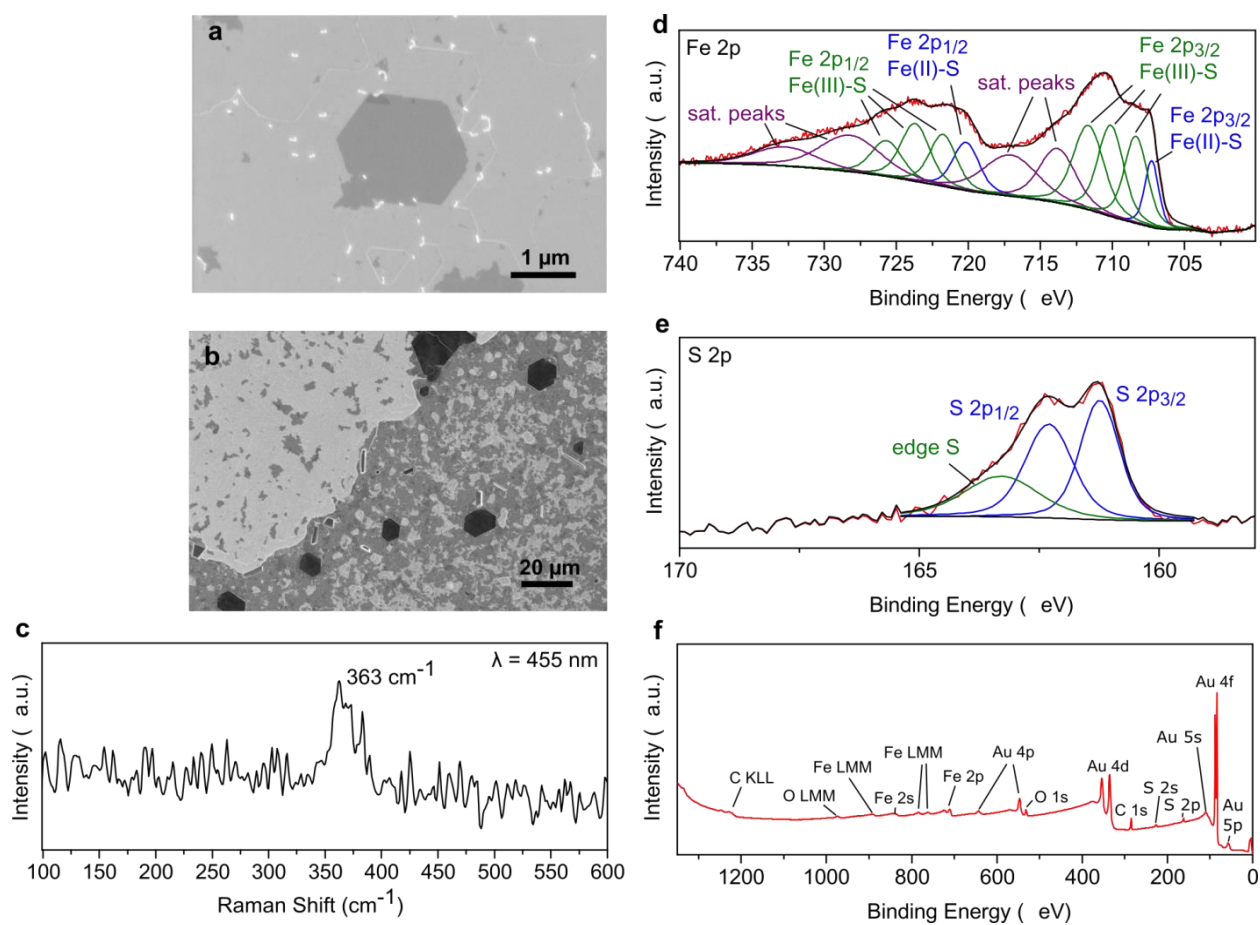

**Supplementary Figure 13. Characterisation of iron sulphide (FeS). (a)** High magnification and **(b)** low magnification SEM images of FeS on gold. **(c)** Raman spectrum of FeS domains on gold. **(d-f)** XPS spectra of the gold surface post-growth: **(d)** Fe 2p, **(e)** S 2p, and **(f)** survey scan.

Iron sulphides show various growth modes on gold as seen in (b), with some regions on the surface appearing to favour the growth of monolayers, while others favouring the growth of multilayers and 3D crystals. The Fe(II)-S peaks at 707.3 eV and 720.8 eV in the Fe 2p XPS spectrum (d) correspond to reported values for mackinawite[20,21], the layered phase of iron sulphide. We attribute the additional peaks at 708.4 eV, 710.1 eV, 711.7 eV, 722.4 eV, 724.1 eV, and 727.2 eV to spin multiplets of $Fe^{3+}$ oxidation states at the edges and defects in the sulphide domains[21]. The peaks at 713.9 eV, 730.5 eV and 716.9 eV, 733.9 eV are attributed to satellite peaks of the $Fe^{2+}$ and $Fe^{3+}$ oxidation states, respectively. The S 2p region shows 3 peaks at 161.6 eV, 162.9 eV and 164.2 eV, which are attributed to the S $2p_{3/2}$, S $2p_{1/2}$ and edge states of the FeS domains [20,22], respectively. A small amount of oxygen can be seen in the survey spectrum; as such, it is plausible that oxide peaks may be convoluted with the satellite peaks in (d). The Raman spectrum (c) shows a peak at 363 cm$^{-1}$ which is associated with the greigite phase of iron sulphide[23] – it is known that the mackinawite phase oxidizes to the greigite phase upon exposure to ambient air[20,24].

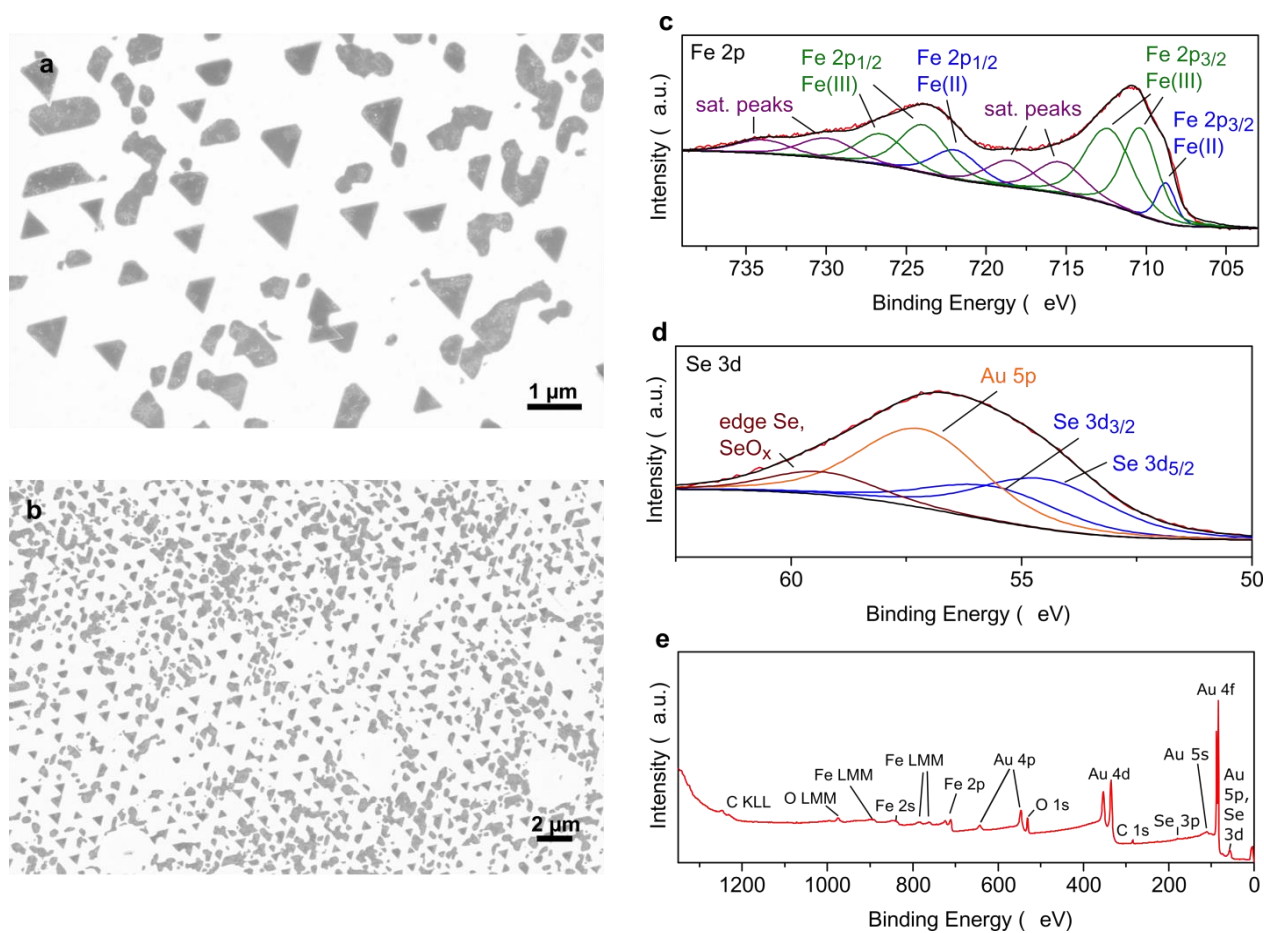

**Supplementary Figure 14. Characterisation of iron selenide (FeSe). (a)** High magnification and **(b)** low magnification SEM images of FeSe on gold. **(c-e)** XPS spectra of the gold surface post-growth: **(c)** Fe 2p, **(d)** Se 3d, and **(e)** survey scan.

Deconvolution of the Fe 2p XPS spectra for FeSe was done using the FeS spectrum as a guide. Since a considerable amount of oxygen is present from the survey spectrum, we have only identified the peaks from the Fe oxidation states, as the iron oxide peaks are convoluted with the selenide peaks. The Fe(II) peaks at 708.8 eV and 722.1 eV correspond to the $Fe^{2+}$ oxidation state. Additional peaks at 710.4 eV, 712.4 eV, 723.9 eV and 726.5 eV are attributed to $Fe^{3+}$ oxidation states. Satellite peaks at 715.4 eV, 729.9 eV and 718.5 eV, 733.8 eV are attributed to the $Fe^{2+}$ and $Fe^{3+}$ states, respectively. Further oxide/multiplet peaks may be convoluted with the satellite and $Fe^{3+}$ multiplet peaks. The peaks at 54.5 eV and 55.6 eV in the Se 3d region correspond to FeSe[25,26].

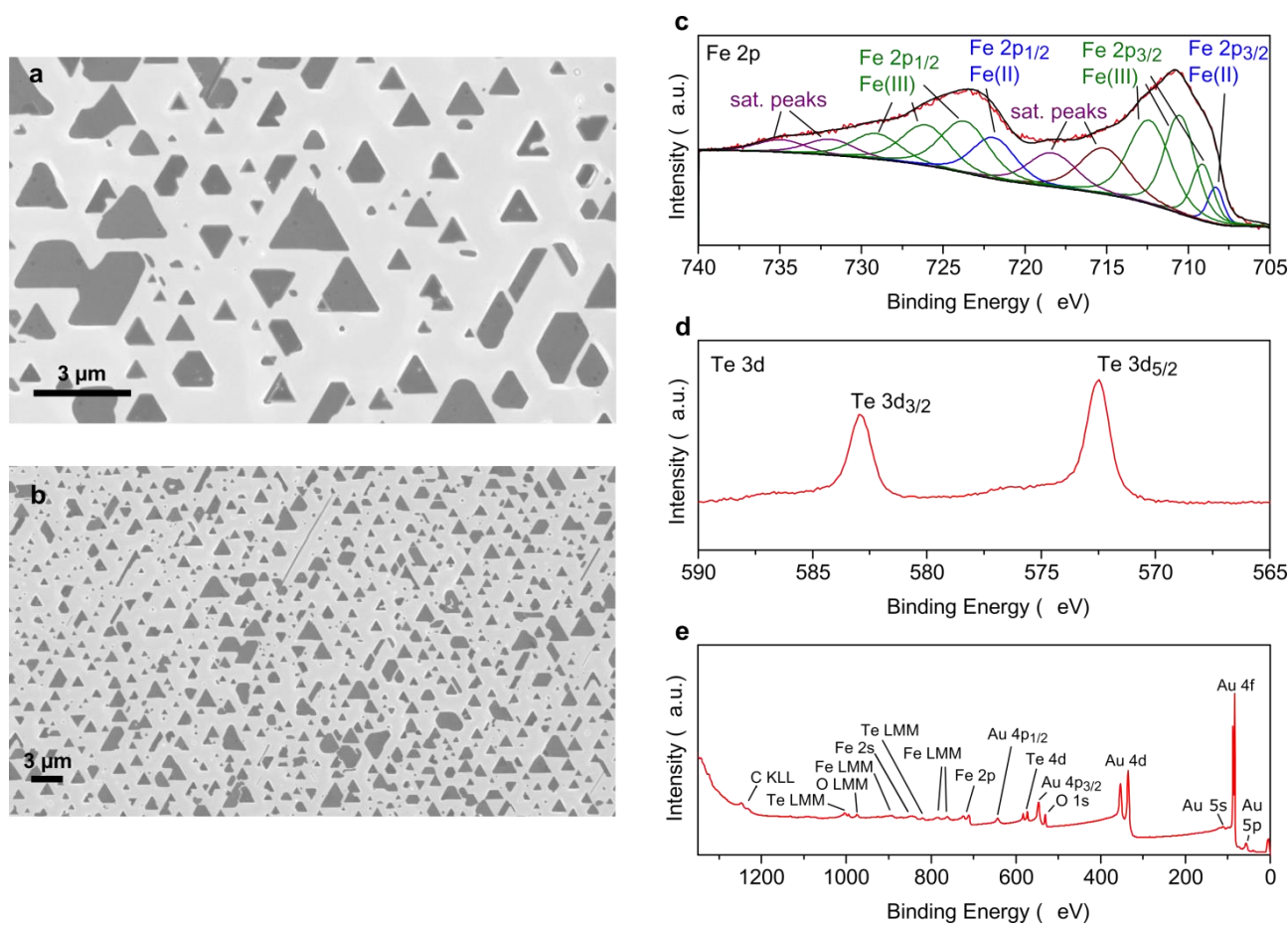

**Supplementary Figure 15. Characterisation of iron telluride (FeTe). (a)** High magnification and **(b)** low magnification SEM images of FeTe on gold. **(c-e)** XPS spectra of the gold surface post-growth: **(c)** Fe 2p, **(d)** Te 3d, and **(e)** survey scan.

Deconvolution of the Fe 2p region for FeTe was done using FeS and FeSe XPS spectra as guides. As with the case of FeSe, oxygen is present on the surface, and as such only the oxidation states are identified in the Fe 2p region. The peaks at 708.3 eV and 721.4 eV are assigned to the $Fe^{2+}$ oxidation state. Further peaks in (c) at 709.1 eV, 710.5 eV, 712.4 eV, 723.4 eV, 726.1 eV and 729.1 eV are attributed to multiplets of the $Fe^{3+}$ oxidation state. The peaks at 715.2 eV, 731.9 eV and 718.3 eV, 734.9 eV are assigned to the $Fe^{2+}$ and $Fe^{3+}$ satellite peaks, respectively. Te 3d peaks at 572.5 eV and 582.9 eV in (d) correspond to FeTe [7,26]. FeTe also shows evidence of epitaxy with the underlying gold substrate, as seen in the SEM images (a-b).

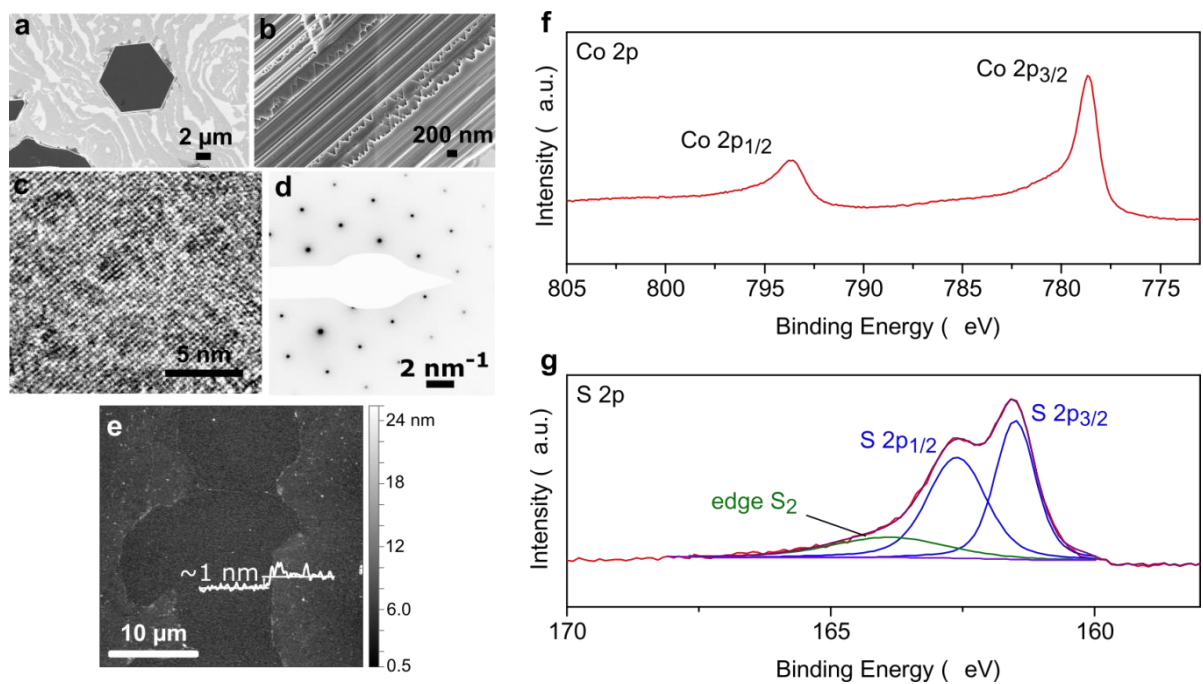

**Supplementary Figure 16. Characterisation of cobalt disulphide (CoS$_2$). (a)** Low magnification SEM image of CoS$_2$ on gold, and **(b)** a high magnification SEM image of one of the edges of the 3D crystals in **(a)**, showing a resemblance to layered crystals. **(c)** high resolution TEM image and (d) SAED pattern of one of the edges of the 3D crystals in (a). Monolayered regions in (a) were too unstable under e-beam irradiation for TEM imaging at 200 kV. (e) AFM of transferred domains in (a) onto 90 nm SiO$_2$/Si substrate, where the measured thickness of the monolayered regions is ~1 nm, not accounting for chemical contrast between tip and surface. **(f-g)** XPS spectra of the gold surface post-growth: **(f)** Co 2p and **(g)** S 2p regions.

The growth behaviour of sulphides of the ferrous metals Co and Fe bear similarities. We observe monolayer films of CoS$_2$ (light gray films in (a)) as well as 3D crystals (black hexagon in (a)) similar to those seen for FeS but not seen for any of the other metals tested in this work. The 3D crystals here appear to resemble layered crystals from the SEM (b), but appear to have a cubic crystal structure as seen from the BF-TEM and SAED patterns in (c-d). The XPS Co 2p peaks at 778.7 eV and 793.7 eV, and the S 2p peaks at 161.5 eV and 162.6 eV correspond to literature values for CoS$_2$[27,28], whereas as the Co:S ratio is 1:2. We attribute the peak at 163.9 eV in the S 2p spectrum to sulphide edges[22]. No Raman signal was measureable on as-grown or transferred films with our setup.

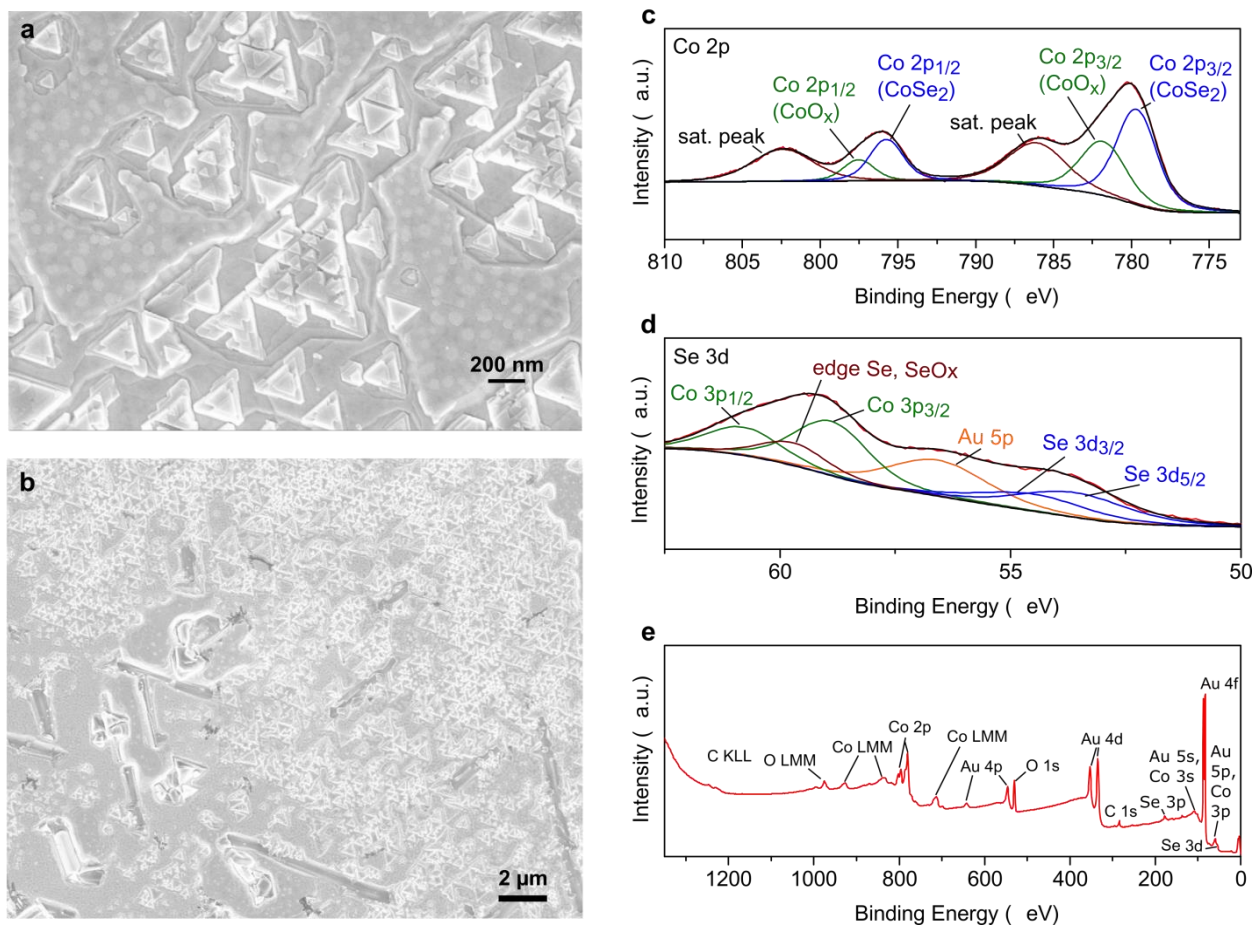

**Supplementary Figure 17. Characterisation of cobalt diselenide (CoSe$_2$).** **(a)** High magnification and **(b)** low magnification SEM images of CoSe$_2$ on gold. **(c-e)** XPS spectra of the gold surface post-growth: **(c)** Co 2p, **(d)** Se 3d, and **(e)** survey scan.

A number of structures are observed in the SEM images (a-b), which we believe are a combination of CoSe$_2$ and cobalt oxides. XPS scan of the Co 2p region confirms the presence of CoSe$_2$: Co 2p peaks at 779.7 eV and 795.8 eV correspond to reported values for CoSe$_2$[29,30]. Co 2p peaks at 781.8 eV and 797.5 eV are attributed to cobalt oxides, while the peaks 786.1 eV and 802.4 eV are satellite peaks of the Co$^{2+}$ oxidation state[29,30]. Se 3d peaks at 53.6 eV and 54.5 eV correspond to CoSe$_2$[29–31].

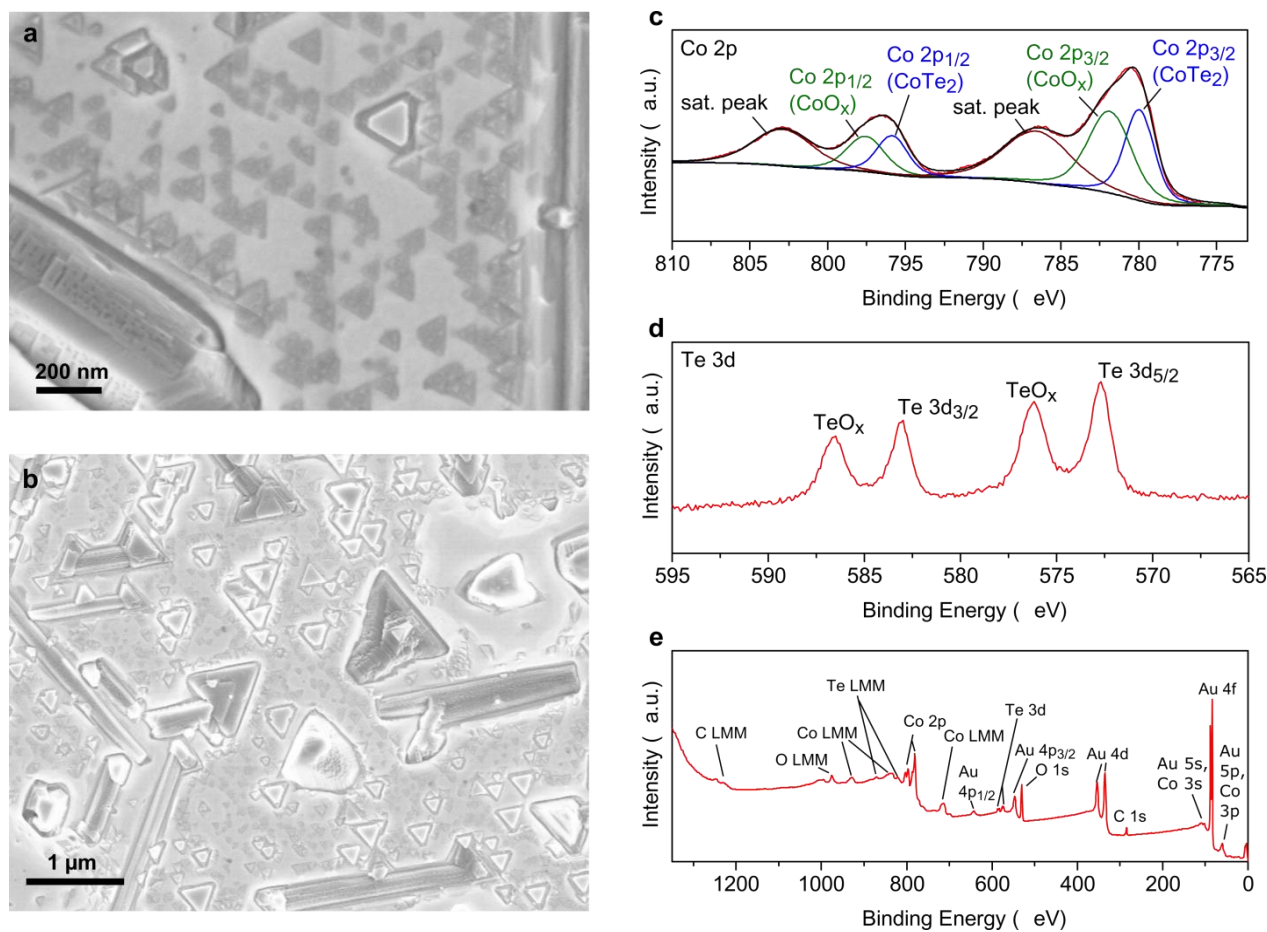

**Supplementary Figure 18. Characterisation of cobalt ditelluride (CoTe$_2$). (a)** High magnification and **(b)** low magnification SEM images of CoTe$_2$ on gold. **(c-e)** XPS spectra of the gold surface post-growth: **(c)** Co 2p, **(d)** Te 3d, and **(e)** survey scan.

A variety of structures are observed in the SEM images (a-b), which bear resemblance to those seen in the case of CoSe$_2$. Deconvolution of the Co 2p spectrum (c) was done using the CoSe$_2$ case as a guide: Co 2p peaks at 779.9 eV and 795.9 eV correspond to CoTe$_2$, while the peaks at 781.9 eV and 797.6 eV are attributed to cobalt oxides; Peaks at 786.6 eV and 802.9 eV are Co$^{2+}$ satellite peaks. Telluride peaks in the Te 3d region are seen at 572.7 eV and 583.0 eV, while the peaks at 576.2 eV and 586.6 eV correspond to oxidised tellurides[7]. CoTe$_2$ appears to be more air-sensitive than CoSe$_2$, as seen from the oxidised telluride peaks and the prominent cobalt oxide peaks, and from higher oxygen content in the survey spectrum. The triangular structures in (a) correspondingly show visible signs of oxidation/degradation.

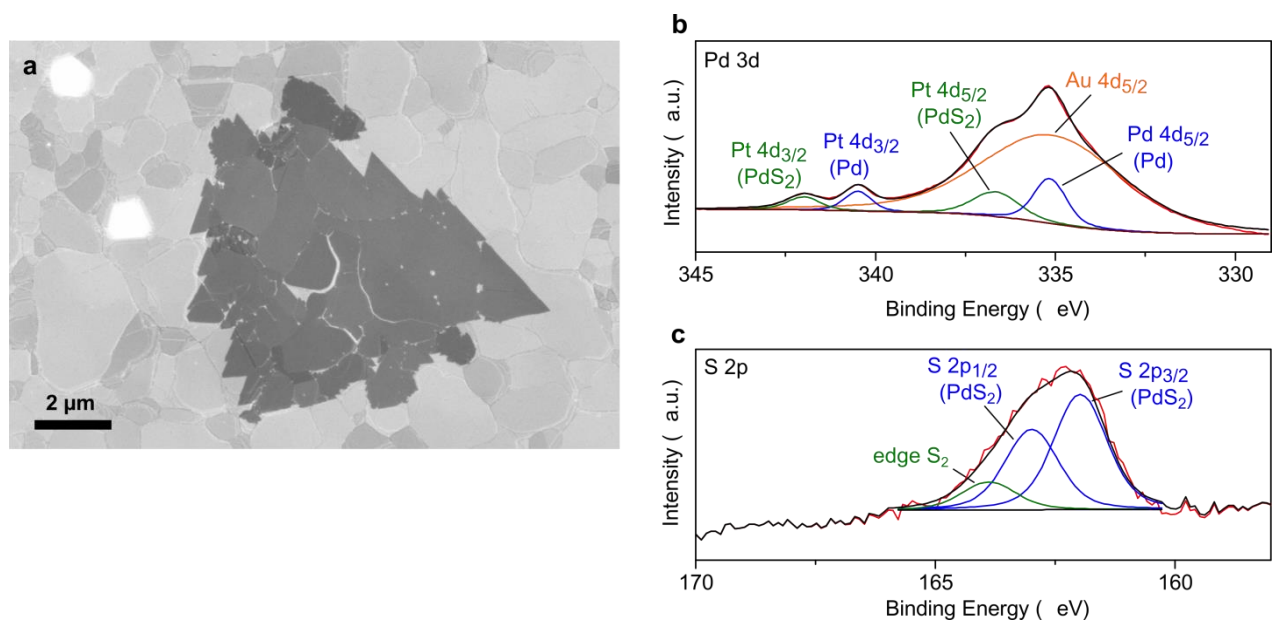

**Supplementary Figure 19. Characterisation of palladium disulphide (PdS$_2$). (a)** SEM image of PdS$_2$ domains on gold. **(b-c)** XPS spectra of the gold surface post-growth: **(b)** Pd 3d and **(c)** S 2p regions. Peaks at 335.2 eV and 340.5 eV in (b) correspond to elemental Pd, while the peaks at 336.7 eV and 342.0 eV correspond to PdS$_2$. The S 2p peaks at 161.9 eV and 163.0 eV in (c) are attributed to PdS$_2$[7], while the peak at 163.9 eV is attributed to edge sulphides[22].

.

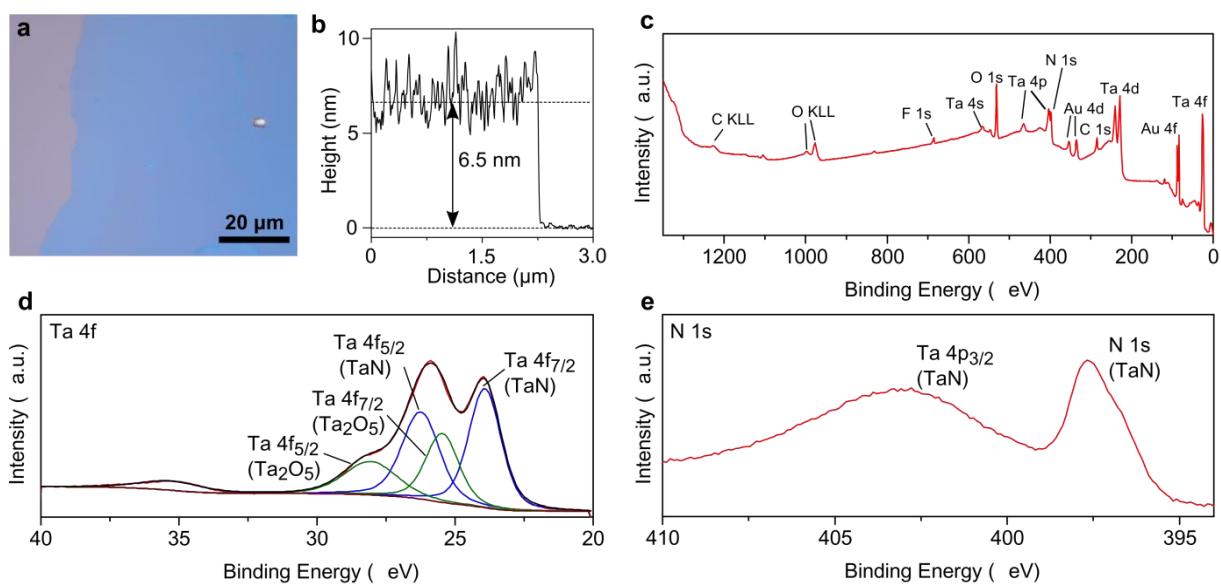

**Supplementary Figure 20. Characterisation of ultrathin tantalum nitride films (TaN).**
**(a)** Optical image of transferred TaN on 90 nm SiO$_2$/Si substrate. **(b)** AFM line profile of transferred TaN films; the film thickness is ~6.5 nm. **(c-e)** XPS spectra of the gold surface post-growth: **(c)** survey spectrum, **(d)** Ta 4f, and **(e)** N 1s scan. Ta 4f$_{7/2}$ and Ta 4f$_{5/2}$ peaks at ~24.0 eV and ~26.0 eV in (d) and the Ta 4p$_{3/2}$ and N 1s peaks in (e) correspond to reported values for hexagonal TaN[32]; the Ta:N ratio here is ~1.1:1. The films show a noticeable amount of oxidation, as seen by the Ta$_2$O$_5$ peaks in (d) and the large oxygen peak in the survey spectrum. The fluorine and carbon signals in the survey spectrum are from backstreamed carbon contaminants[33].

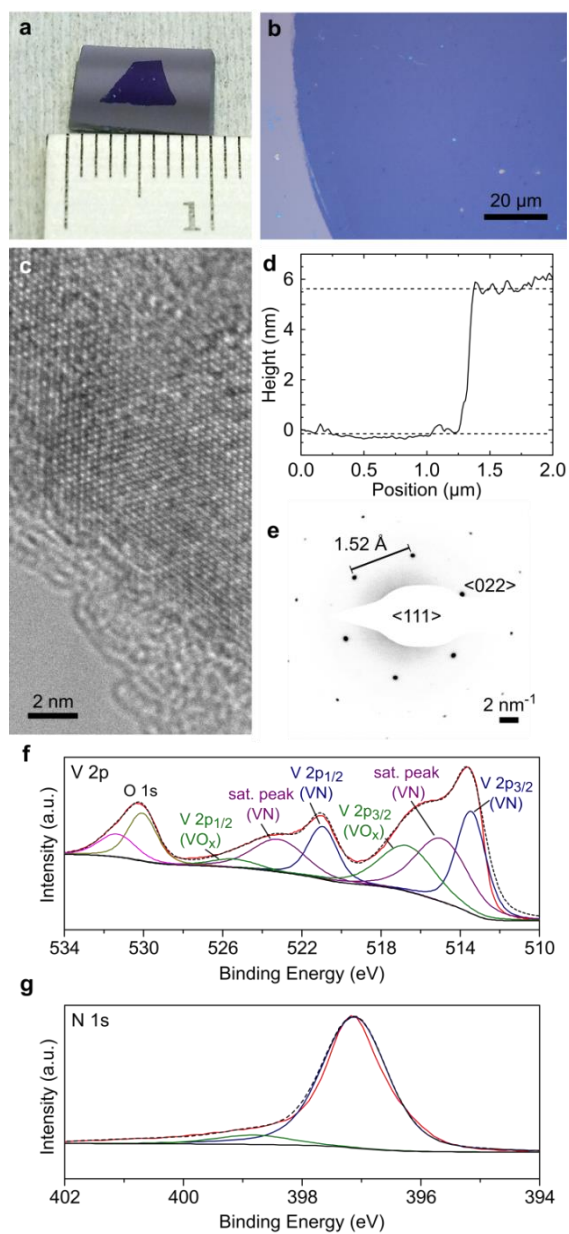

**Supplementary Figure 21. Characterisation of ultrathin vanadium nitride films (VN).**
**(a)** Photograph of ultrathin vanadium nitride film transferred to 90 nm $SiO_2$ on Si substrate (scale in cm). **(b)** Optical microscope image of film in (a). **(c)** High resolution bright field TEM micrograph of suspended vanadium nitride film. **(d)** Atomic force microscopy line scan of edge of transferred vanadium nitride film; the measured AFM thickness is ~ 5.5 nm. **(e)** SAED pattern of suspended vanadium nitride film. **(f-g)** XPS spectra of as-grown vanadium nitride on gold. Partial oxidation is visible in the spectrum in (f).

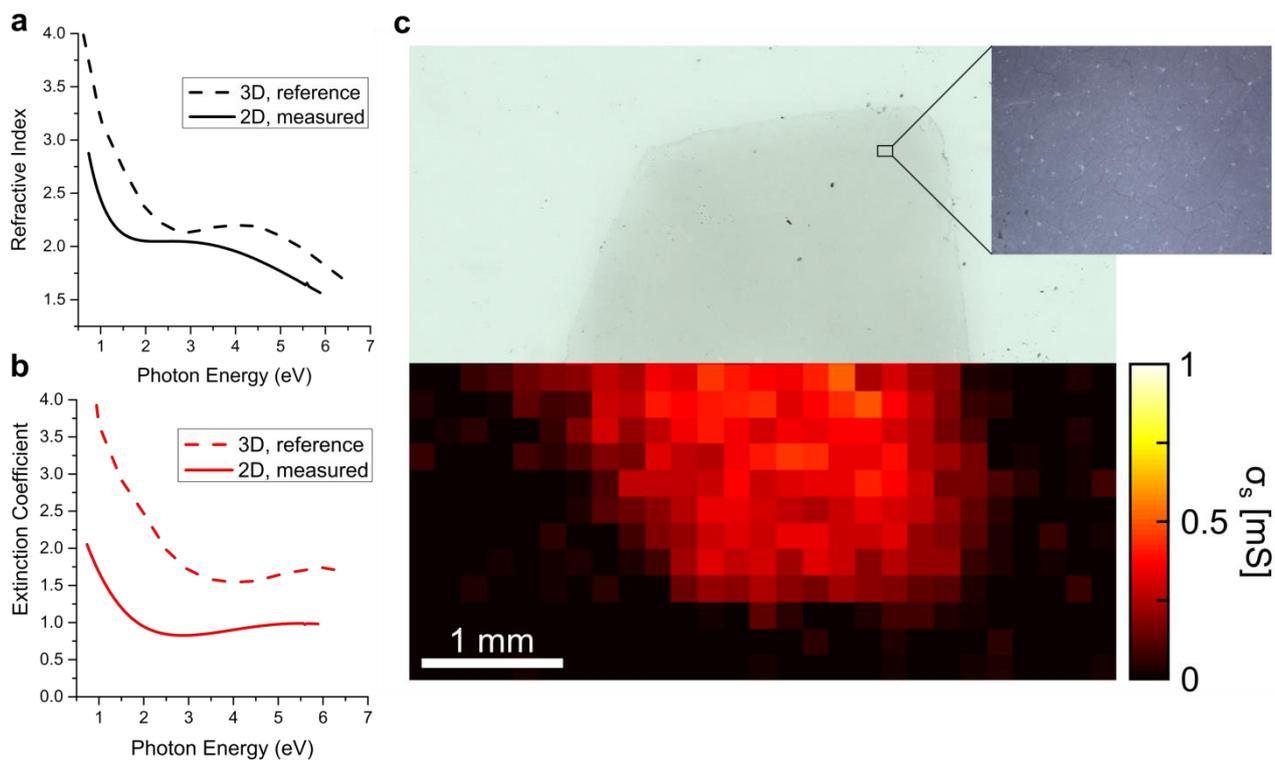

**Supplementary Figure 22. Ellipsometry, optical and THz-TDS characterisation of ultrathin vanadium nitride films on quartz. (a-b)** Optical functions of 5.5 nm VN (labelled 2D) extracted by least squares fitting of ellipsometry spectra using a parameterised optical function model. The model consists of one Drude oscillator (three fitting parameters) to model free carrier absorption in the infrared, and one Lorentz oscillator (three fitting parameters) to model the response in the visible and near ultraviolet. Optical functions obtained from literature[34] for the bulk VN films are also shown (labelled 3D). **(c)** Optical/THz-TDS sheet conductivity map (mirrored) of transferred VN films on quartz. The sheet conductivity map shows the average sheet conductivity in the range from 0.8-0.9 THz.

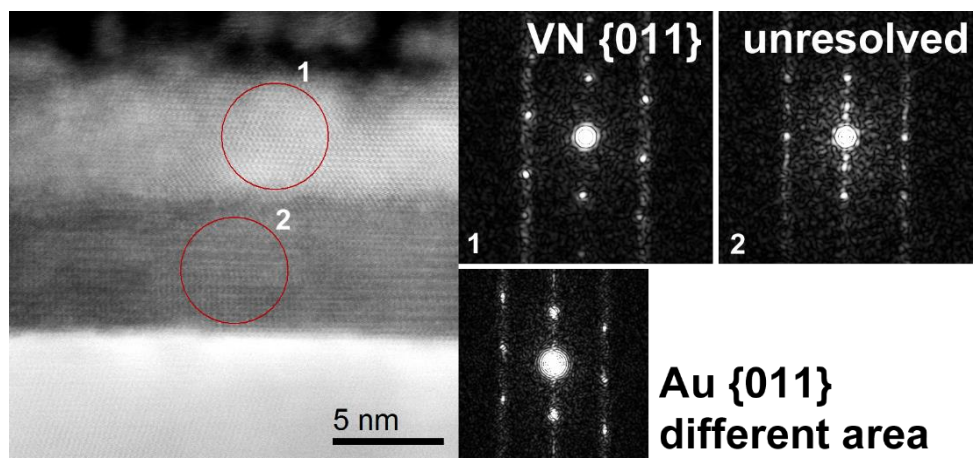

**Supplementary Figure 23. STEM characterisation of as-grown VN films on gold.** The images show the dark-field STEM image of the FIB cross-section sample. The bottom high contrast film is Au, as confirmed by indexing the FFT as shown on the right (the FFT was taken from a separate area). Two layers can be identified on top of Au, marked by 1 and 2. FFT of layer 1 matches with the Fm-3m structure of cubic VN in the {011} orientation. Layer 2 shows a darker contrast. The FFT could not be matched with known vanadium nitride, vanadium oxide, or Au-V alloy structures, but the structure is clearly different from the VN layer in 1. The FFT of layer 2 was indexed against the following compounds (ICSD code): 644862, 64701, 8236, 60486, 34033, 66665, 94768, 612459, 58547, 58611, 61246.

Supplementary Figure 21a-b show a centimetre-scale continuous layer of ultrathin vanadium nitride (VN) transferred onto a 90 nm $SiO_2$ on Si substrate. The size of the film here was limited only by the dimensions of the growth substrate. The thickness of the film is ~5.6 nm, as measured by AFM (Supplementary Fig. 21d), and is uniform across the sample, as confirmed by optical microscopy and ellipsometry (Supplementary Fig. 21b, Supplementary Fig. 22a-b). HRTEM and SAED of the VN films (Supplementary Fig. 21c, e) establishes an FCC lattice structure with the {111} facet perpendicular to the growth substrate. Notably, the lattice spacing here, 1.52 Å, is ~4% larger than that of bulk cubic VN, 1.46 Å (ICDD database). XPS characterisation of as-grown VN films on gold (Supplementary Fig. 21f, g) shows the expected nitride bonding between vanadium and nitrogen[35], with a V:N ratio of ~1:1. Some degree of oxidation was always present in the films even when grown under highly reducing conditions, likely due to exposure of the films to ambient conditions. The sheet resistance of transferred VN layers as determined by van der Pauw measurements is ≈ 2 kΩ/□ immediately after transfer, increasing to ≈ 50 kΩ/□ after 100 days of exposure to ambient conditions, which supports the idea that VN layers are air- or humidity-sensitive. Terahertz time domain spectroscopy results performed 15 days after transfer show an average sheet resistivity of 2 kΩ/□ across the transferred VN film (Supplementary Fig. 22c).

We also performed cross-sectional TEM measurements of as-grown VN films on gold (Supplementary Fig. 23). In these measurements, we observe the previously described ultrathin cubic VN phase, and an unexpected second phase beneath which does not appear to survive the transfer process. The crystal structure differs notably from the transferrable layer above, but we were unable to unequivocally determine the structure by convergent beam electron diffraction, scanning TEM or high resolution TEM.

# References

1. Buron, J. D. *et al.* Electrically Continuous Graphene from Single Crystal Copper Verified by Terahertz Conductance Spectroscopy and Micro Four-Point Probe. *Nano Lett.* **14,** 6348–6355 (2014).
2. Buron, J. D. *et al.* Graphene Conductance Uniformity Mapping. *Nano Lett.* **12,** 5074–5081 (2012).
3. Shearer, C. J., Slattery, A. D., Stapleton, A. J., Shapter, J. G. & Gibson, C. T. Accurate thickness measurement of graphene. *Nanotechnology* **27,** 125704 (2016).
4. Godin, K., Cupo, C. & Yang, E.-H. Reduction in step height variation and correcting contrast inversion in dynamic AFM of $WS_2$ monolayers. *Sci. Rep.* **7,** 17798 (2017).
5. Gutiérrez, H. R. *et al.* Extraordinary Room-Temperature Photoluminescence in Triangular $WS_2$ Monolayers. *Nano Lett.* **13,** 3447–3454 (2013).
6. Elías, A. L. *et al.* Controlled synthesis and transfer of large-area $WS_2$ sheets: from single layer to few layers. *ACS Nano* **7,** 5235–5242 (2013).
7. Moulder, J. F. & Chastain, J. *Handbook of X-ray photoelectron spectroscopy: a reference book of standard spectra for identification and interpretation of XPS data*. (Physical Electronics Division, Perkin-Elmer Corporation, 1992).
8. Yim, C. *et al.* High-performance hybrid electronic devices from layered $PtSe_2$ films grown at low temperature. *ACS Nano* **10,** 9550–9558 (2016).
9. Tonndorf, P. *et al.* Photoluminescence emission and Raman response of monolayer $MoS_2$, $MoSe_2$, and $WSe_2$. *Opt. Express* **21,** 4908–4912 (2013).
10. Jiang, Y. C., Gao, J. & Wang, L. Raman fingerprint for semi-metal $WTe_2$ evolving from bulk to monolayer. *Sci. Rep.* **6,** 19624 (2016).
11. Lee, C.-H. *et al.* Tungsten Ditelluride: a layered semimetal. *Sci. Rep.* **5,** 10013 (2015).
12. Lu, X. *et al.* Large-area synthesis of monolayer and few-layer $MoSe_2$ films on $SiO_2$ substrates. *Nano Lett.* **14,** 2419–2425 (2014).
13. Abdallah, W. A. & Nelson, A. E. Characterization of $MoSe_2$(0001) and ion-sputtered $MoSe_2$ by XPS. *J. Mater. Sci.* **40,** 2679–2681 (2005).
14. Nam, D., Lee, J.-U. & Cheong, H. Excitation energy dependent Raman spectrum of $MoSe_2$. *Sci. Rep.* **5,** 17113 (2015).
15. Han, G. H. *et al.* Absorption dichroism of monolayer 1T′-$MoTe_2$ in visible range. *2D Mater.* **3,** 31010 (2016).
16. Zhao, S. *et al.* Two-dimensional metallic $NbS_2$: growth, optical identification and transport properties. *2D Mater.* **3,** 25027 (2016).
17. Kanazawa, T. *et al.* Few-layer $HfS_2$ transistors. *Sci. Rep.* **6,** 22277 (2016).
18. Yuan, J. *et al.* Facile synthesis of single crystal vanadium disulfide nanosheets by chemical vapor deposition for efficient hydrogen evolution reaction. *Adv. Mater.* **27,**

5605–5609 (2015).
19. Rout, C. S. *et al.* Synthesis and characterization of patronite form of vanadium sulfide on graphitic layer. *J. Am. Chem. Soc.* **135,** 8720–8725 (2013).
20. Boursiquot, S., Mullet, M., Abdelmoula, M., Génin, J.-M. & Ehrhardt, J.-J. The dry oxidation of tetragonal $FeS_{1-x}$ mackinawite. *Phys. Chem. Miner.* **28,** 600–611 (2001).
21. Han, D. S., Batchelor, B. & Abdel-Wahab, A. XPS analysis of sorption of selenium(IV) and selenium(VI) to mackinawite (FeS). *Environ. Prog. Sustain. Energy* **32,** 84–93 (2013).
22. Bruix, A. *et al.* In Situ Detection of Active Edge Sites in Single-Layer $MoS_2$ Catalysts. *ACS Nano* **9,** 9322–9330 (2015).
23. Bourdoiseau, J.-A., Jeannin, M., Rémazeilles, C., Sabot, R. & Refait, P. The transformation of mackinawite into greigite studied by Raman spectroscopy. *J. Raman Spectrosc.* **42,** 496–504 (2011).
24. Lennie, A. R. & Vaughan, D. J. Spectroscopic studies of iron sulfide formation and phase relations at low temperatures. *Miner. Spectrosc. a Tribut. to Roger G. Burn.* **5,** 117–131 (1996).
25. Chen, T. K. *et al.* Low-temperature fabrication of superconducting FeSe thin films by pulsed laser deposition. *Thin Solid Films* **519,** 1540–1545 (2010).
26. Telesca, D., Nie, Y., Budnick, J. I., Wells, B. O. & Sinkovic, B. Impact of valence states on the superconductivity of iron telluride and iron selenide films with incorporated oxygen. *Phys. Rev. B* **85,** 214517 (2012).
27. Zhang, H. *et al.* A metallic $CoS_2$ nanopyramid array grown on 3D carbon fiber paper as an excellent electrocatalyst for hydrogen evolution. *J. Mater. Chem. A* **3,** 6306–6310 (2015).
28. Chen, C.-J. *et al.* An integrated cobalt disulfide ($CoS_2$) co-catalyst passivation layer on silicon microwires for photoelectrochemical hydrogen evolution. *J. Mater. Chem. A* **3,** 23466–23476 (2015).
29. Kwak, I. H. *et al.* $CoSe_2$ and $NiSe_2$ Nanocrystals as Superior Bifunctional Catalysts for Electrochemical and Photoelectrochemical Water Splitting. *ACS Appl. Mater. Interfaces* **8,** 5327–5334 (2016).
30. Li, H. *et al.* Template synthesis of $CoSe_2/Co_3Se_4$ nanotubes: tuning of their crystal structures for photovoltaics and hydrogen evolution in alkaline medium. *J. Mater. Chem. A* **5,** 4513–4526 (2017).
31. McCarthy, C. L., Downes, C. A., Schueller, E. C., Abuyen, K. & Brutchey, R. L. Method for the Solution Deposition of Phase-Pure $CoSe_2$ as an Efficient Hydrogen Evolution Reaction Electrocatalyst. *ACS Energy Lett.* **1,** 607–611 (2016).
32. Lehn, J.-S. M., van der Heide, P., Wang, Y., Suh, S. & Hoffman, D. M. A new precursor for the chemical vapor deposition of tantalum nitride films. *J. Mater. Chem.* **14,** 3239–3245 (2004).
33. Shivayogimath, A. *et al.* Probing the gas-phase dynamics of graphene chemical vapour deposition using in-situ UV absorption spectroscopy. *Sci. Rep.* **7,** 6183 (2017).
34. Pflüger, J., Fink, J., Weber, W., Bohnen, K. P. & Crecelius, G. Dielectric properties of $TiC_x$, $TiN_x$, $VC_x$, and $VN_x$ from 1.5 to 40 eV determined by electron-energy-loss spectroscopy. *Phys. Rev. B* **30,** 1155–1163 (1984).
35. Glaser, A. *et al.* Oxidation of vanadium nitride and titanium nitride coatings. *Surf. Sci.* **601,** 1153–1159 (2007).